\documentstyle[12pt,a41,epsfig]{article}

\newcommand{\Li}{{\rm Li}}
\newcommand{\Sf}{{\rm S}_{1,2}}
\newcommand{\CLra}{{C_{L,q}^{N(1)}\over C_{L,g}^{N(1)}}}
\newcommand{\bCLra}{{\bar C_{L,q}^{N(1)}\over \bar C_{L,g}^{N(1)}}}
\newcommand{\CLrb}{{C_{L,q}^{N(2)} \over C_{L,g}^{N(1)}}}
\newcommand{\CLrc}{{C_{L,g}^{N(2)} \over C_{L,g}^{N(1)}}}
\newcommand{\Gqqa}{\gamma_{qq}^{N(0)}}
\newcommand{\Gqga}{\gamma_{qg}^{N(0)}}
\newcommand{\Ggqa}{\gamma_{gq}^{N(0)}}
\newcommand{\Ggga}{\gamma_{gg}^{N(0)}}
\newcommand{\Gqqb}{\gamma_{qq}^{N(1)}}
\newcommand{\Gqgb}{\gamma_{qg}^{N(1)}}
\newcommand{\Ggqb}{\gamma_{gq}^{N(1)}}
\newcommand{\Gggb}{\gamma_{gg}^{N(1)}}

\newcommand{\CLga}{C_{L,g}^{N(1)}}
\newcommand{\CLqb}{C_{L,q}^{N(2)}}
\newcommand{\CLgb}{C_{L,g}^{N(2)}}
\newcommand{\Cq}{C_{2,q}^{N(1)}}
\newcommand{\Cg}{C_{2,g}^{N(1)}}
\newcommand{\bb}{{\beta_1 \over \beta_0}}
\newcommand{\ZqqNT}{Z_{qq}^{(T)N(1)}}
\newcommand{\ZqgNT}{Z_{qg}^{(T)N(1)}}
\newcommand{\ZgqNT}{Z_{gq}^{(T)N(1)}}
\newcommand{\ZggNT}{Z_{gg}^{(T)N(1)}}

\begin{document}
\sloppy
\thispagestyle{empty}
\begin{flushleft}
DESY 98--144    \hfill
{\tt hep-ph/0004172}\\
March 2000    \\
\end{flushleft}

\mbox{}
\vspace*{\fill}
\begin{center}
 
{\LARGE\bf On the Drell--Levy--Yan Relation to {\boldmath $O(\alpha_s^2)$}}

\vspace{4em}
\large
J. Bl\"umlein$^a$, V. Ravindran$^{a,b}$ and W.L. van Neerven$^{a,c}$
\normalsize
\\
\vspace{4em}
{\it$^a$DESY Zeuthen}\\
{\it    Platanenallee 6, D--15735 Zeuthen, Germany}\\

\vspace{4mm}
{\it $^b$Mehta Research Institute of Mathematics and Mathematical Physics
(MRI)}\\
{\it Chhatnag Road, Jhusi, Allahabad, 211019, India}\\

\vspace{4mm}
{\it 
$^c$Instituut-Lorentz,
Universiteit Leiden,}\\ {\it
P.O. Box 9506, 2300 HA Leiden, The Netherlands.}
\end{center}
\vspace*{\fill}
\begin{abstract}
\noindent
We study the validity of a relation by Drell, Levy and Yan (DLY) 
connecting the deep inelastic structure (DIS) functions and the 
single-particle fragmentation functions in $e^+e^-$ annihilation which 
are defined in the spacelike ($q^2<0$) and timelike ($q^2>0$) regions 
respectively. Here $q$ denotes the momentum of the virtual photon 
exchanged in the deep inelastic scattering process or the annihilation
process. An extension of the DLY-relation, which originally was only
derived in the scaling parton model, to all orders in QCD leads to a 
connection between the two evolution kernels determining the 
$q^2$-dependence of the DIS structure functions and the fragmentation 
functions respectively. In relation to this we derive the transformation 
relations between the space--and time--like splitting functions up to 
next-to-leading order (NLO) and the coefficient functions up to NNLO both
for unpolarized and polarized scattering. It is shown that the evolution 
kernels describing the combined singlet evolution for the structure 
functions $F_2(x,Q^2)$, $F_L(x,Q^2)$ where $Q^2=|q^2|$ or $F_2(x,Q^2), 
\partial F_2(x,Q^2)/\partial \ln(Q^2)$ and the corresponding fragmentation 
functions satisfy the DLY relation up to next-to-leading order. We also 
comment on a relation proposed by Gribov and Lipatov.
\end{abstract}
\vspace*{\fill}
\newpage
\section{Introduction}

\vspace{1mm}
\noindent
Neutral current deep inelastic lepton--nucleon scattering and single 
nucleon inclusive production in $e^+ e^-$ pair--annihilation are formally 
related by crossing the kinematic channels. Already before the advent of 
Quantum Chromodynamics (QCD) Drell, Levy, and Yan mentioned the 
possibility~\cite{BAS1,BAS2} that the deep inelastic scattering structure
functions at the one side and the nucleon fragmentation functions 
in $e^+ e^-$ pair--annihilation on the 
other side may be related by an analytic continuation from the $t$-- to 
the $s$--channel. The hadronic tensors for the space--like process of 
deep inelastic scattering (DIS) and the time--like single nucleon inclusive
reaction would therefore be related by
\begin{equation}
W_{\mu\nu}^{(S)}(q,p) = - W_{\mu\nu}^{(T)}(q,-p)~.
\label{eqDLY}
\end{equation}
Here $p$ denotes the nucleon momentum and $q$ is the 4--momentum transfer
to the hadronic system, with $q^2 < 0$ for deep inelastic scattering and 
$q^2 > 0$ for $e^+ e^-$--annihilation.

At that time the  physical nucleons were considered as bound states built up 
of bare nucleons and pions in the context of the Yukawa theory. The 
interactions were described by Bethe--Salpeter~\cite{BS} or 
Faddeev--type~\cite{FAD} equations and their 
generalization~\cite{DL}~\footnote{Later on massive vector meson
ladder--models were studied in \cite{gl1,SULF}, where the crossing 
relation Eq.~(\ref{eqDLY}) was verified for the respective kernels.}
aiming at a perturbative description of the structure and fragmentation
functions. In these theories neither infra--red nor collinear 
singularities are occurring. One may think off a general representation of
the structure and fragmentation functions in terms of current--current 
expectation values. However it was already shown \cite{BAS1}
that for $e^+ e^-$-annihilation also diagrams of distinct connectedness
appear (cf. also~\cite{old1}) which are absent in DIS so that a proof
of Eq.~(\ref{eqDLY}) at the non--perturbative level becomes very difficult. 
The relation could be established for the 
aforementioned ladder--models~\cite{BAS1,old1,old2}~\footnote{For a
review on the early developments see~\cite{Men}.}. 
Within QCD the picture
changes. Here it turns out that a thorough perturbative description of the 
structure functions and fragmentation functions is not possible. However, 
perturbation theory may be used to describe the scaling violations of 
these functions at large values of $|q^2|$ where the running coupling constant
is small. The QCD--improved parton 
model, moreover, exhibits a similarity with the approach by Drell, Levy, 
and Yan, since in the range where single parton states
 are dominating, i.e. for
the contributions of lowest twist, the non--perturbative contributions 
factorize.  One therefore may calculate the respective one--particle
evolution kernels and study their behavior under the crossing from the 
$t$-- to the $s$--channel. A further complication in the case of a 
vector--theory as QCD is the emergence of infrared and also collinear
singularities which have an essential impact on the crossing because of 
the behavior of the kernels at $x=1$~\footnote{Eq.~(\ref{eqDLY}) was
originally postulated assuming that those terms are absent~\cite{BAS1}.
See also the subsequent discussion in \cite{lan1} pointing out that the 
exponent of the structure functions $\propto (1-x)^p$ near 
$x=1$~\cite{DYBG} needs not to be integer.}. Here $x$ denotes the Bjorken 
scaling variable which will be defined differently for the timelike and 
spacelike region except for $x=1$ where both definitions lead to the same 
value for $x$. As a consequence of the 
Bloch--Nordsiek theorem~\cite{BN} the kernels become distribution--valued
for $x=1$~\cite{GS}. 

In leading order for unpolarized scattering the crossing relations mentioned 
above were given in~\cite{LIP}. Similar relations hold in 
the polarized case. In this order these kernels are nothing but the lowest 
order splitting functions $P_{kl}^{(0)}(x)$, which are obtained from the 
inverse Mellin transforms of the anomalous dimensions
$\gamma_{kl}^{N(0)}$ \cite{LO}.

It is the aim of the present paper to investigate the validity
of the Drell--Levy--Yan (DLY) relation, if applied to perturbatively calculable
partonic structure functions and quantities related to them up to
the level of two--loop order. To establish this crossing relation between
space-- and time--like processes one has to study scheme--invariant
quantities which are the physical evolution kernels for specific choices
of observables as the unpolarized and polarized structure and 
fragmentation functions or derivatives of them w.r.t. $q^2$. 
Furthermore, conditions are derived for the transformation of the
splitting and coefficient functions from the space-- to the time--like 
case. For the coefficient functions we extend the discussion to the NNLO
level. Other relations between the splitting functions
such as supersymmetric relations and relations due to conformal symmetry
were discussed elsewhere, cf. e.g.~\cite{brv,bms}.

The paper is organized as follows. Basic relations for the deep 
inelastic structure and fragmentation functions are summarized in 
section~2. In section~3 scheme--invariant combinations of coefficient
and splitting functions are constructed for the space-- and time--like
processes both for unpolarized and polarized deep inelastic reactions where
we consider two principal examples. The Drell--Levy--Yan relation is
studied in detail in section~4. We also comment on a relation by
Gribov and Lipatov~\cite{GL} which emerged in the same context.  
Section~5 contains the conclusions. In the appendix we present the differences
between to the space-- and time--like coefficient functions at 
$O(\alpha_s^2)$ as well as the convolution relations which are needed for 
the investigation of the DLY--relation.
\section{Structure Functions and Fragmentation Functions}

\vspace{1mm}
\noindent
Deep inelastic scattering (DIS) of a lepton ($l$) off a hadron target 
($P$) is described by the process
\begin{equation}
\label{DIS}
l(k_1)+P(p) ~\rightarrow~ l(k_2) + `X' \quad , \quad q=k_1-k_2 \quad ,
\quad q^2=-Q^2<0\,.
\end{equation}
where $`X'$ represents an inclusive final state. When a 
single gauge boson is exchanged between the incoming lepton and the 
hadron  the above process factorizes into the leptonic part and the 
remaining hadronic part. In the case of forward scattering the scattering     
matrix element can be written in terms of the leptonic tensor 
$L_{\mu\nu}$ and the hadronic tensor $W_{\mu\nu}$ by
\begin{equation}
\label{MAT}
|{\boldmath M}|^2 = L^{\mu\nu} W_{\mu\nu}\,.
\end{equation}
The hadronic tensor~\cite{SF} contains the  unpolarized and polarized
deep inelastic structure functions $F_i$ and $g_i$.
If the process is mediated by photon only we have $i=1,2$ in both the
polarized and unpolarized case. Notice that instead of $F_1$ we can also
take the longitudinal structure function $F_L$. 
At asymptotic values of the kinematic   variables structure functions only
depend on $Q^2$ and the Bjorken scaling variable 
\begin{equation}
\label{XB}
x_B = \frac{Q^2}{2 p.q},~~~~~~~0 \leq x_B \leq 1~.
\end{equation}
In QCD the $Q^2$ dependence of the structure functions       is only
logarithmic and it accounts for the violation of scaling. In the context
of the parton model the structure functions can be 
expressed in terms of quark and gluon densities and the 
corresponding spacelike coefficient functions $C_{i,k}^{(S)}$ ($k=q,g$)
~\footnote{Similar 
relations hold for the polarized structure functions $g_1(x,Q^2)$ and
$g_2(x,Q^2)$ on the level of twist~2.}
\begin{eqnarray}
\label{eqDIS}
F_i^{(S)}(x_B,Q^2)&=&  x_B            
\sum_{j=1}^{N_f} e_j^2 \int_{x_B}^1
{dz \over z} \Bigg [ { 1 \over N_f}\,f_q^S\left({x_B \over z},\mu_f^2\right) 
C_{i,q}^{(S)S}\left(z,{Q^2 \over \mu_f^2}\right) 
\!+\!f_g\left({x_B \over z},\mu_f^2\right) 
\nonumber\\[2ex]
&&~~~~~~~~~~
\times C_{i,g}^{(S)}\left(z,{Q^2 \over \mu_f^2}\right)
+f_{q_j}^{NS}\left({x_B \over z},\mu_f^2\right) 
C_{i,q}^{(S)NS}\left(z,{Q^2 \over \mu_f^2}\right) \Bigg ]~,
\label{eqFiS}
\nonumber\\[2ex]
i=2,L\,,&&
\end{eqnarray}
where $e_j$ denotes the charge of the $j$th quark flavor and $N_f$ represents
the number of light flavors. The scale $\mu_f$, appearing in 
the above equation, denotes the factorization scale which is introduced 
while removing the collinear singularities from the partonic structure
functions.   
In addition one encounters a dependence on the renormalization scale 
$\mu_r$ which arises in the renormalization procedure. For convenience this 
scale is put equal to the factorization scale in the following. 
Notice that the structure 
functions $F_i$ and $g_i$ do not depend on these scales. However 
the parton densities and the coefficient functions, which do depend on these 
scales, satisfy renormalization group equations which will be shown below.

In Eq.~(\ref{eqFiS}) the index $(S)$ in the structure functions indicates
the space--like nature of the process ($q^2 < 0$). Furthermore in Eq.
(\ref{eqDIS}) appear the singlet `S' and non-singlet `NS' combinations of
parton densities which are defined by
\begin{eqnarray}
\label{eqS}
f_q^S\left(z,\mu_f^2\right)&=&\sum_{i=1}^{N_f}
\Bigg[f_{q_i}\left(z,\mu_f^2\right)+
f_{\bar q_i}\left(z,\mu_f^2\right)\Bigg]\,,
\end{eqnarray}
and
\begin{eqnarray}
\label{eqNS}
f_{q_i}^{NS}\left(z,\mu_f^2\right)&=&
f_{q_i}\left(z,\mu_f^2\right)+
f_{\bar q_i}\left(z,\mu_f^2\right)
-{1 \over N_f} f_q^S\left(z,\mu_f^2\right)\,,
\end{eqnarray}
respectively.
Corresponding formulae hold for polarized scattering. In this case
the polarized parton densities and polarized coefficient functions are
denoted by $\Delta f_k(z,\mu_f^2)$ ($k=q,g$) and $\Delta
C_{i,k}(z,Q^2/\mu_f^2)$
($i=1,2$).

Whereas in deep inelastic scattering the constituent structure of the 
nucleons is studied, hadroproduction at $e^+e^-$ colliders provides us with
information about the fragmentation process of these constituents into
the hadrons. This information is contained in the fragmentation functions
observed in the reaction \cite{BAS1}  
\begin{eqnarray}
\label{FRAG}
l(k_1) + \bar l(k_2) \rightarrow \bar P(p) + `X'\quad , \quad q=k_1+k_2 \quad ,
\quad q^2 \equiv Q^2>0 \,,
\end{eqnarray}
where the symbols have he same meaning as in Eq. (\ref{DIS}).
These fragmentation functions are the analogues of the DIS structure functions 
Therefore in the QCD improved parton model these functions can be expressed in 
a similar way in terms of parton fragmentation densities $D_k$ ($k=q,g$)
multiplied by timelike coefficient functions i.e.
\begin{eqnarray}
\label{eqFRA}
F_i^{(T)}(x_E,Q^2)&=&x_E  \sum_{j=1}^{n_f} e_j^2  \int_{x_E}^1
{dz \over z} \Bigg [ {1 \over N_f }\,D_q^S\left({x_E \over z},\mu_f^2\right) 
C_{i,q}^{(T)S}\left(z,{Q^2 \over \mu_f^2}\right) 
\!+\!D_g\left({x_E \over z},\mu_f^2\right) 
\nonumber\\[2ex]
&&~~~~~~~~~~
\times C_{i,g}^{(T)}\left(z,{Q^2 \over \mu_f^2}\right)
+D_{q_j}^{NS}\left({x_E \over z},\mu_f^2\right) 
C_{i,q}^{(T)NS}\left(z,{Q^2 \over \mu_f^2}\right) \Bigg ],
\nonumber\\[2ex]
i=2,L\,, &&
\end{eqnarray}
where the corresponding scaling variable for the process in Eq.~(9)
is defined by
\begin{equation}
\label{XE}
x_E = \frac{2 p.q}{Q^2},~~~~~~~0 \leq x_E \leq 1~,
\end{equation}
The symbol $T$ appearing within parentheses in Eq. (\ref{eqFRA})
denotes that the fragmentation functions
are measured in time--like processes. The scales $\mu_f$ and $\mu_r$
are defined in the same way as in Eq. (\ref{eqDIS}) where like in DIS we set
the renormalization scale equal to the factorization scale. Furthermore
the definitions for the singlet and non-singlet parton fragmentation functions 
are the same as those for the parton densities given in 
Eqs.~(\ref{eqS}, \ref{eqNS}). Similarly as in DIS one can also study
the annihilation processes in Eq. (\ref{FRAG}) where the hadron $P$ is 
polarized. This entails the definition of the polarized fragmentation
functions denoted by $g_1^{(T)}$ and $g_2^{(T)}$
for which one can present a similar formula as in Eq. (\ref{eqFRA}).
Very often one also encounters the transverse
structure function which in the timelike and spacelike case is given by
\begin{eqnarray}
\label{eqF1}
F_1^{(R)}(x,Q^2)&=&{1 \over 2 x} \Bigg [F^{(R)}_2(x,Q^2)
     -F^{(R)}_L(x,Q^2) \Bigg] \,, 
\nonumber\\[2ex]
\mbox{with} \qquad R=S \,\, (x=x_B) && R=T \,\, (x=x_E)\,.
\end{eqnarray}

\section{Scheme--invariant Combinations}

\vspace{1mm}
\noindent
In this section we give a short outline of the origin of the factorization 
scheme dependence of the anomalous dimensions (splitting functions) and
the coefficient functions. We also show how this dependence disappears in
the evolution of the structure functions w.r.t. the kinematic   variable
$Q^2$. The discussion below deals with the DIS structure functions but
the conclusions also hold for the fragmentation functions.
The partonic structure functions denoted by $\hat {\cal F}_{i,k}$ ($i=1,2,L$,
$k=q,g$), representing the QCD radiative corrections, contain various 
divergences. First these divergences 
have to be regularized for which the most convenient way is to choose
the method of $n$--dimensional 
regularization. Using this method the singularities
reveal themselves in the form of pole terms of the type $1/\epsilon^j$,
with $n=4+\epsilon$, in the quantity $\hat {\cal F}_{i,k}$.
The infrared divergences cancel between virtual and 
bremsstrahlung contributions by virtue of the Bloch--Nordsieck 
theorem~\cite{BN}. Due to the Kinoshita-Lee-Nauenberg theorem~\cite{kln} 
all the final state mass singularities are canceled too since the DIS structure
function is an inclusive quantity. Then one is left with
only two types of singularities. The first one originates from the 
ultraviolet region. This type of singularities is removed via a redefinition 
of the parameters appearing in the QCD Lagrangian. An example is the coupling 
constant which becomes equal to $\alpha_s(\mu_r^2)$ where $\mu_r$ is the 
renormalization scale. After coupling constant renormalization the hadronic 
structure function can be written as follows
\begin{equation}
\label{PAR}
F_i(x,Q^2)=\sum_{k=q,g} \left (
{\cal F}_{ik} \Big(\alpha_s(\mu_r^2),{Q^2 \over \mu^2},{\mu^2 \over \mu_r^2},
\epsilon \Big )
\otimes \hat f_k\right )(x)\,,
\end{equation}
where the symbol $\otimes$ denotes the Mellin--convolution defined by
\begin{equation}
\label{CONV}
(f\otimes g)(z) = \int_0^1 dz_1 \int_0^1 dz_2 f(z_1) g(z_2)
\delta(z-z_1 z_2)\,.
\end{equation}
Furthermore $\hat f_k$ is defined as the bare parton density which is scale 
independent and is an unphysical object because of the singular behavior
of ${\cal F}_{ik}$. Notice that the latter depends on the scale $\mu_r$
and therefore on the renormalization scheme w.r.t. the coupling constant.
The parameter $\mu$ originates from $n$--dimensional 
regularization because
in this method the coupling constant gets a dimension.
The second type of singularity originates from the collinear region which can 
be attributed to the vanishing mass of the initial state parton represented 
by either the (anti-) quark or the gluon. Hence the $\epsilon$ in 
Eq. (\ref{PAR}) represents the collinear
singularities which are removed from the partonic structure function 
via mass factorization and transferred to a transition function
$\Gamma_{lk}$ as follows
\begin{equation}
\label{FACT}
\hat {\cal F}_{ik}(z,\alpha_s(\mu_r^2),{Q^2 \over \mu^2},{\mu^2 \over \mu_r^2},
\epsilon) =\sum_{l=q,g} \left (
C_{i,l} \Big(\alpha_s(\mu_r^2),{Q^2 \over \mu_f^2},
{\mu_f^2 \over \mu_r^2}\Big )
\otimes
\Gamma_{lk}\Big(\alpha_s(\mu_r^2), {\mu_f^2 \over \mu^2},
{\mu_f^2 \over \mu_r^2}, \epsilon\Big)\right )(z) \,.
\end{equation}
This procedure provides us with the coefficient function denoted by
$C_{i,l}$. Substitution of Eq. (\ref{FACT}) into Eq. (\ref{PAR}) leads to
the result
\begin{equation}
\label{eqFA}
F_i(x,Q^2)=\sum_{l=q,g} \left (
C_{i,l} \Big(\alpha_s(\mu_r^2),{Q^2 \over \mu_f^2},
{\mu_f^2 \over \mu_r^2}\Big ) \otimes 
f_l\Big(\alpha_s(\mu_r^2),{\mu_f^2 \over \mu^2},{\mu_f^2 \over \mu_r^2}\Big)
\right )(x)\,,
\end{equation}
where the renormalized parton density is defined as
\begin{equation}
\label{DENS}
f_l\Big(z,\alpha_s(\mu_r^2),{\mu_f^2 \over \mu^2},{\mu_f^2 \over \mu_r^2}\Big)
=\sum_{k = q,g}\left (
\Gamma_{lk}\Big(\alpha_s(\mu_r^2), {\mu_f^2 \over \mu^2}
,{\mu_f^2 \over \mu_r^2}, \epsilon\Big ) \otimes \hat f_k \right )(z)~.
\end{equation}
Since the mass factorization can be carried out in various ways one is
left with an additional scheme dependence which comes on top of the
renormalization scheme dependence entering the coupling constant
in Eq. (\ref{PAR}). The former only shows up in the
parton densities and the coefficient functions and it only disappears in
specific combinations representing physical quantities. Hence physical
quantities are invariant under scheme transformation. Like in the case of
renormalization,  mass factorization leads to the introduction of a
scale $\mu_f$ called mass factorization scale which is related to the
factorization scheme dependence. Like in the latter case $\mu_f$ drops
out in physical quantities as the DIS structure functions 
or fragmentation functions.
The change of the parton densities and the coefficient functions
with respect to a variation in the scales $\mu_r$ and $\mu_f$
is determined by the renormalization group equation (RGE) \cite{callan}
The renormalization group equation of the parton densities follow from
the one presented for the transition functions $\Gamma_{lk}$. The latter
takes the following form 
\begin{eqnarray}
\label{EVOL1}
&& \left (\Big [ \Big \{\mu_f^2 \frac{\partial}{\partial \mu_f^2} 
+  \beta(a_s(\mu_f^2)) 
\frac{\partial}{\partial a_s(\mu_f^2)}\Big \}{\bf 1}\delta_{lm}  - 
{1 \over 2} P_{lm}(a_s(\mu_f^2),\epsilon) \Big ]
\otimes\Gamma_{mk}\Big(a_s(\mu_f^2), {\mu_f^2 \over \mu^2},1,\epsilon\Big )
\right )(z) =0\,,
\nonumber\\[2ex]
&& a_s(\mu_f^2) \equiv \frac{\alpha_s(\mu_f^2)}{4\pi} \quad , \quad 
{\bf 1} =\delta(1-z)\,,
\end{eqnarray}
where we have set $\mu_r=\mu_f$ for simplicity. The functions 
$P_{ij}\Big(a_s,\epsilon,z\Big)$ appearing in the above equation are
the splitting functions. Furthermore the beta-function is defined by
\begin{equation}
\label{BETA}
\mu_r^2 {d~a_s(\mu_r^2) \over d~ \mu_r^2 } = - \beta_0 a_s^2(\mu_r^2)
-  \beta_1 a_s^3(\mu_r^2) \cdots \,,
\end{equation}
The same equation as in Eq. (\ref{EVOL1}) also applies to the parton density 
because of the definition
in Eq. (\ref{DENS}). The scale dependence of the coefficient function
in  Eq. (\ref{eqFA}) is given by
\begin{eqnarray}
\label{EVOL2}
\left (\Big [ \Big \{\mu_f^2 \frac{\partial}{\partial \mu_f^2}
+  \beta(a_s(\mu_f^2))
\frac{\partial}{\partial a_s(\mu_f^2)}\Big \}{\bf 1}\delta_{lm}  +
{1 \over 2} P_{lm}(a_s(\mu_f^2),\epsilon) \Big ]
\otimes C_{i,m}\Big(a_s(\mu_f^2), {Q^2 \over \mu_f^2},1\Big )\right )(z)
=0\,.
\nonumber\\
\end{eqnarray}
As has been mentioned above scheme transformations such as
\begin{equation}
\Gamma_{lk} \rightarrow \sum_{m=q,g}Z_{lm} \otimes \bar \Gamma_{mk}\quad , 
\quad  C_{i,l} \rightarrow \sum_{m=q,g}\bar C_{i,m} \otimes Z^{-1}_{ml} \,,
\end{equation}
will not alter the physical observable like e.g. 
the structure functions and
fragmentation functions. The relation between the splitting 
and coefficient functions computed in two different schemes is found to
be
\begin{eqnarray}
\label{eqS1}
P_{lk}&=&\sum_{\{m,n\}=q,g}Z_{lm} \otimes \bar P_{mn}
\otimes (Z^{-1})_{nk} -2 \beta(a_s)\sum_{m=q,g}  Z_{lm}
\otimes {d \over d a_s}  (Z^{-1})_{mk}\,,
\\[2ex]
\label{eqS2}
C_{i,l}&=&\sum_{m=q,g}\bar C_{i,m} \otimes (Z^{-1})_{ml}~.
\end{eqnarray}
Below we present the relation between the coefficient functions computed 
in two different schemes up to order $a_s^2(Q^2)$. 
Notice that we have chosen here
$\mu_f^2=Q^2$ in order to get rid off the logarithms $\ln(Q^2/\mu_f^2)$
which usually appear. Up to $O(a_s^2)$ one obtains
\begin{eqnarray}
C_{i,q}&=&\delta(1-z) + a_s \bigg(\bar C_{i,q}^{(1)} + Z_{qq}^{(1)}\bigg)
+a_s^2 \bigg( \bar C_{i,q}^{(2)} +Z_{qq}^{(2)} +(Z_{qq}^{(1)})^2
+Z_{qg}^{(1)} \otimes Z_{gq}^{(1)}
\nonumber\\[2ex]
&&+\bar C_{i,q}^{(1)} \otimes Z_{qq}^{(1)} +\bar  C_{i,g}^{(1)} \otimes
 Z_{gq}^{(1)} \bigg) +\cdots \,,
\\[2ex]
C_{i,g}&=&a_s \bigg(\bar C_{i,g}^{(1)} +Z_{qg}^{(1)}\bigg) + 
a_s^2 \bigg( \bar C_{i,g}^{(2)} +Z_{qg}^{(2)} + Z_{qg}^{(1)} \otimes 
\big(Z_{gg}^{(1)}+Z_{qq}^{(1)}\big)
\nonumber\\[2ex]
&&+\bar C_{i,q}^{(1)} \otimes Z_{qg}^{(1)} + \bar C_{i,g}^{(1)} \otimes 
Z_{gg}^{(1)} \bigg) + \cdots\,. 
\end{eqnarray}
In the subsequent part of the paper it is much more convenient to derive
the expressions in the Mellin transform representation so that one can
avoid the convolution symbol $\otimes$. The Mellin transform of a function
$f(z)$ is given by
\begin{equation}
\label{Mel1}
f^{(N)}= \int_0^1 dz~z^{N-1}~f(z)
\end{equation}
In this way Eq. (\ref{CONV}) can be written as
\begin{equation}
\label{Mel2}
(f\otimes g)^N= \int_0^1 dz~z^{N-1}~(f\otimes g)(z)= f^N \cdot
g^N~.
\end{equation}
Since the structure functions are scheme independent they become 
renormalization group invariants. Hence they satisfy the RG equation
\begin{equation}
\label{EVOL3}
\Bigg [ \mu_f^2 \frac{\partial}{\partial \mu_f^2}
+  \beta(a_s(\mu_f^2))
\frac{\partial}{\partial a_s(\mu_f^2)} \Bigg ] F_i^N(x,Q^2) =0
\end{equation}
The equation above follows from combining Eqs. (\ref{EVOL1},
\ref{EVOL2}) and (\ref{eqFA}). However the independence of
the structure function on the scales $\mu_f$ and $\mu_r$ is not manifest
when multiplying parton densities with coefficient functions.
In particular when the perturbation series of a physical quantity is
computed up to finite order there is a residual dependence on these
unphysical scales (see e.g. \cite{rmq}. Their influence is expected to
become smaller when higher order terms in the perturbation series are
taken into account \cite{NV}. 
To avoid the problem of the factorization scheme dependence of the structure
function when the perturbation series is truncated up to fixed order
it is better to study evolution equations for the structure functions
with respect to a {\it physical} scale which is represented by
a kinematic   variable like $Q^2$. In these type of evolution equations
the kernels are factorization scheme independent order by order in perturbation
theory. However the dependence on the choice of renormalization scheme
and therefore the dependence on $\mu_r$ remains so that one is able
to obtain a better estimate of the theoretical error on $\alpha_s$.
For such an equation one needs two different structure functions
called $F_A(x,Q^2)$ and $F_B(x,Q^2)$. Examples are $A=2$ and $B=L$
or $F_A$ and $Q^2 d~F_A/d~Q^2$. Limiting ourselves to the singlet case,
the evolution equation for the non-singlet structure functions is even more 
simple, one can write
\begin{eqnarray}
\label{eqFA1}
&& F_I^N(Q^2)=f_q^N\left (a_s(\mu_f^2),{\mu_f^2 \over Q_0^2 }\right ) 
C_{I,q}^N\left (a_s(\mu_f^2),{Q^2 \over \mu_f^2 }\right )
+ f_g^N\left (a_s(\mu_f^2),{\mu_f^2 \over Q_0^2 }\right )
C_{I,g}^N\left (a_s(\mu_f^2),{Q^2 \over \mu_f^2 }\right )\,,
\nonumber\\[2ex]
&& I=A,B\,,
\end{eqnarray}
Here one can view the $C_{I,l}^N$ ($I=A,B$, $l=q,g$) as matrix elements
so that the equation above has the form
\begin{equation}
\label{eqevo1}
\left( \begin{array}{c}
F^N_A\\ {F^N_B}
\end{array} \right)= 
\left( \begin{array} {cc}
C^N_{Aq} & C^N_{Ag}\\
C^N_{Bq} & C^N_{Bg}
\end{array} \right)
\left( \begin{array}{c}
f_q^N\\ {f_g^N}
\end{array} \right)~.
\end{equation}
The coefficient functions satisfy the RG-equation in Eq. (\ref{EVOL2})
and the solution is given by the T-ordered exponential \cite{bur}
\begin{eqnarray}
\label{SOL1}
C_{I,l}^N\left ( a_s(\mu_f^2),\frac{Q^2}{\mu_f^2} \right )=
C_{I,m}^N \left ( a_s(Q^2),1 \right ) \left ( T_{a_s} \left [exp 
\left \{-\int_{a_s(\mu_f^2)}^{a_s(Q^2)} dx\, \frac{\gamma^N(x)}{2\beta(x)} 
\right \} \right ]\right )_{ml}\,,
\end{eqnarray}
where $\gamma^N$ is the anomalous dimension matrix defined by
\begin{equation}
\gamma_{lk}^N = -\int_0^1 dz z^{N-1} P_{lk}(z)\,.
\end{equation}
We will now differentiate the coefficient functions w.r.t. $Q^2$
\begin{eqnarray}
\label{SOL2}
&& Q^2 \frac{\partial C_{I,k}^N ( a_s(\mu_f^2),Q^2/\mu_f^2 ) }
{\partial Q^2}=\beta (a_s(Q^2)) 
\frac{\partial C^N_{I,k} ( a_s(\mu_f^2),Q^2/\mu_f^2 )}{\partial a_s(Q^2)}=
\nonumber\\[2ex]
&& \Bigg [ \beta (a_s(Q^2) \frac{\partial C^N_{I,m}(a_s(Q^2),1) }
{\partial a_s(Q^2)} \left (C^N\right )_{m,J}^{-1}(a_s(Q^2),1) 
\nonumber\\[2ex]
&& - {1 \over 2}C^N_{I,m}(a_s(Q^2),1) \gamma^N_{mn}(a_s(Q^2))
\left (C^N \right )_{n,J}^{-1}(a_s(Q^2),1)\Bigg ] 
C^N_{J,k} ( a_s(\mu_f^2),\frac{Q^2}{\mu_f^2} )\,.
\end{eqnarray}
One can     show that the expression above is invariant under scheme
transformations. The latter are given by
\begin{eqnarray}
\label{eqS3}
\gamma^N_{lk}&=&\sum_{\{m,n\}=q,g}Z^N_{lm} \bar \gamma^N_{mn}
\left (Z^N \right )_{nk}^{-1} +2 \beta(a_s)\sum_{m=q,g}  Z^N_{lm}
{\partial \over \partial a_s}  \left (Z^N \right )_{mk}^{-1}\,,
\\[2ex]
\label{eqS4}
C^N_{I,l}&=&\sum_{m=q,g}\bar C^N_{I,m}\left (Z^N \right )_{ml}^{-1},
\qquad
\left (C^N \right )_{l,I}^{-1}=\sum_{m=q,g}Z^N_{lm} 
\left( \bar C^N \right )_{m,I}^{-1}\,.
\end{eqnarray}
Since the $Q^2$-dependence only resides in the coefficient function
the same evolution equation as in Eq. (\ref{SOL2}) also applies
to $F^N_I$ in Eq. (\ref{eqFA1}).
For a short-hand notation we introduce the evolution variable $t$
\begin{equation}
t=-{2 \over \beta_0} \ln\left({a_s(Q^2) \over a_s(Q_0^2)}\right),
\end{equation}
so that we obtain
\begin{equation}
\label{eqevo2}
{\partial \over \partial t}\left( \begin{array}{c}
F^N_A\\
{F^N_B}
\end{array} \right)= -{1 \over 4}
\left( \begin{array} {cc}
K^N_{AA} & K^N_{AB}\\
K^N_{BA} & K^N_{BB}
\end{array} \right)
\left( \begin{array}{c}
F^N_A\\
{F^N_B}
\end{array} \right)~,
\end{equation}
where the physical (scheme invariant) kernel is given by
\begin{eqnarray}
\label{KERN1}
K^N_{IJ}=\Bigg [ -4 \frac{\partial C^N_{I,m}(t) }
{\partial t } \left (C^N\right )_{m,J}^{-1}(t)
 - {\beta_0 a_s(Q^2) \over  \beta(a_s(Q^2))}
C^N_{I,m}(t) \gamma^N_{mn}(t)
\left (C^N \right )_{n,J}^{-1}(t)\Bigg ]\,.
\end{eqnarray}
The kernels $K^N_{IJ}$ depend both on the anomalous dimensions
$\gamma^N_{lk}$ and the coefficient functions $C^N_{I,l}(Q^2)$ but the latter
two quantities are combined in a factorization scheme independent way. This
factorization scheme independence of $K^N_{IJ}$ holds order by order in 
perturbation theory.
Using the series expansions for the anomalous dimensions and coefficient 
functions in terms of the strong coupling constant
\begin{equation}
\gamma^N_{lk} = \sum_{n=0}^{\infty} a_s^{n+1}(Q^2) 
\left (\gamma^N\right )_{lk}^{(n)} 
\quad ,\quad C^N_{I,l}(Q^2) = \sum_{n=0}^{\infty} a_s^n(Q^2) 
\left (C^N\right )_{I,l}^{(n)} 
\quad , \quad l,k=q,g \quad , \quad I=A,B\,,
\end{equation}
one can compute order by order the coefficients in the perturbation series of 
the kernel
\begin{eqnarray}
\label{KERN2}
K^N_{IJ} = \sum_{n=0}^{\infty} 
a_s^n(Q^2) \left (K^N \right )_{IJ}^{(n)}~.
\end{eqnarray}
Notice that the coefficients $\left (K^N \right )_{IJ}^{(n)}$
are not invariant with respect to a
finite renormalization of the coupling constant. This dependence is
removed when the perturbation series in Eq. (\ref{KERN2}) is resummed in all 
orders.
\subsection{\boldmath{$F_2(x,Q^2)$} and $F_L(x,Q^2)$}

\vspace{1mm}
\noindent
Let us consider now two specific examples, choosing
the structure functions $F_2(x,Q^2)$ and $F_L(x,Q^2)$ or
the structure function $F_2(x,Q^2)$ and its slope $\partial F_2(x,Q^2)
/\partial t$ as the observables $F_{A,B}(x,Q^2)$.
in this combination of observables it is convenient to normalize
the structure function $F_L(x,Q^2)$ to its gluonic contribution in lowest 
order. This is because $F_L(x,Q^2)$ vanishes in zeroth order of $\alpha_s$
due to the Callan--Gross relation, cf.~Eq.~(\ref{eqF1}),
contrary to the 
structure function $F_2(x,Q^2)$. Therefore this normalization accounts for 
keeping the same order in the coupling constant for the two quantities
\begin{equation}
F^N_A(Q^2)=F^{N(S)}_2(Q^2),\quad \quad   F^N_B(Q^2)=\displaystyle 
{F^N_L(Q^2) 
\over a_s(Q^2) C_{L,g}^{N(1)} }~.
\end{equation}
Since both the coefficient functions $C_{Lq}^{(1)}$ and $C_{Lg}^{(1)}$
are scheme invariants such a normalization is possible. We expand now
the kernels $K_{IJ}^N$ for this choice of observables into a
series in $a_s$. The lowest order contribution is
well-known,~cf.~e.g.~\cite{CAT},
\begin{equation}
\label{eqG11}
\begin{array}{rclcrcl}
K_{22}^{N(0)}\!\!&=&\!\!\displaystyle{\Gqqa- \CLra \Gqga } &
 \hspace{.5cm}&
K_{2L}^{N(0)}\!\!&=&\!\!\displaystyle{\Gqga} 
\nonumber\\[2ex] 
\\
K_{L2}^{N(0)}\!\!&=&\!\!\displaystyle{\Ggqa-\left( \CLra \right)^2 
\Gqga} & \hspace{.5cm}& 
K_{LL}^{N(0)}\!\!&=&\!\!\displaystyle{\Ggga+\CLra \Gqga} 
\nonumber\\[2ex]
&&+ \displaystyle{ \CLra \Big(\Gqqa-\Ggga\Big)}~. 
\\
\nonumber
\end{array}
\end{equation}
To next-to-leading order in $a_s(Q^2)$, one finds
\begin{eqnarray}
\label{eqG12}
K_{22}^{N(1)}&=&\Gqqb-\bb \Gqqa-\CLra \left(\Gqgb-\bb \Gqga\right) 
+\CLra \Cg \Gqqa 
\nonumber\\[2ex]
&&-\left[\CLrb+\left(\CLra\right)^2 \Cg-\CLra \CLrc\right] \Gqga
+\Cg \Ggqa 
\nonumber\\[2ex]
&& -\CLra \Cg \Ggga + 2 \beta_0 \left(\Cq-\CLra \Cg\right)  
\\[2ex]
K_{2L}^{N(1)}&=&\Gqgb-\bb \Gqga-\Cg (\Gqqa-\Ggga)+2 \beta_0 \Cg
\nonumber\\[2ex]
&&+\left(\Cq+\CLra \Cg-\CLrc\right) \Gqga
\\[2ex]
K_{L2}^{N(1)}&=&\Ggqb-\bb \Ggqa+\CLra\left(\Gqqb-\bb\Gqqa \right)
\nonumber\\[2ex] 
&&-\left( \CLra \right)^2 \left(\Gqgb-\bb\Gqga\right)-\CLra 
\left(\Gggb-\bb\Ggga\right)
\nonumber\\[2ex] 
&&+\left[\CLrb -\CLra \Cq+\left(\CLra\right)^2 \Cg \right] 
\Gqqa
\nonumber\\[2ex] 
&&-\left[\left(\CLra \right)^3 \Cg+2 \CLra \CLrb - 
\left(\CLra\right)^2 \CLrc \right .
\nonumber\\[2ex]
&&\left. -\left( \CLra\right)^2 \Cq \right] \Gqga
+\left( \CLra \Cg -\Cq +\CLrc \right) \Ggqa
\nonumber\\[2ex]
&&-\left[ \CLrb +\left(\CLra\right)^2 \Cg -\CLra \Cq \right] \Ggga
\nonumber\\[2ex] 
&&+2 \beta_0 \left( \CLrb- \CLra \CLrc \right)
\\[2ex]
\label{eqG13}
K_{LL}^{N(1)}&=& \Gggb-\bb \Ggga+\CLra \left(\Gqgb-\bb \Gqga\right)
\nonumber\\[2ex]
&&-\CLra \Cg \Gqqa +\left[\CLrb-\CLra \CLrc +\left(\CLra\right)^2 
\Cg\right]\Gqga
\nonumber\\[2ex]
&&-\Cg \Ggqa + \CLra \Cg  \Ggga +2 \beta_0 \CLrc 
\end{eqnarray}
It is evident that the representation in terms of Mellin--moments is
of advantage when compared to the corresponding $x$--space 
expressions. 
In the latter case one has to find the inverse Mellin transforms of
quantities where $C_{i,k}^N$ and $\gamma_{kl}^N$ appear in the denominators
of the expressions above which in general is not possible.
\subsection{\boldmath{$F_2(x,Q^2)$ and $\partial F_2(x,Q^2)/\partial t$}}

\vspace{1mm}
\noindent
A second example concerns the structure function $F_2(x,Q^2)$ and its
slope. Both quantities are well measurable in the present--day deep
inelastic scattering experiments. The observables $F_{A,B}(x,Q^2)$ are
here
\begin{equation}
F_A^N(Q^2)=F^{(S)N}_2(Q^2),\quad \quad F_B^N(Q^2)={\partial \over 
\partial t} F^{(S)N}_2(Q^2) 
\end{equation}
This example has been considered before in \cite{scheme}.
In leading order one obtains
\begin{equation}
\label{eqG21}
\begin{array}{rclcrcl}
K_{22}^{N(0)}&=&0 & \hspace{.5cm}& 
K_{2d}^{N(0)}&=&-4 \nonumber \nonumber \\ \\
K_{d2}^{N(0)}&=&\displaystyle{{1 \over 4} \Bigg(\Gqqa \Ggga-\Gqga 
\Ggqa\Bigg)} & \hspace{.5cm}&
K_{dd}^{N(0)}&=&\displaystyle{\Gqqa+\Ggga }~.\nonumber
\nonumber 
\end{array}
\end{equation}
The next-to-leading order kernels read~:
\begin{eqnarray}
\label{eqG22}
K_{22}^{N(1)}&=&0
\\[2ex]
K_{2d}^{N(1)}&=&0
\\[2ex]
K_{d2}^{N(1)}&=&{1 \over 4} \Bigg[\Ggga \Gqqb+\Gggb \Gqqa -\Gqgb 
\Ggqa -\Gqga \Ggqb\Bigg]
\nonumber\\[2ex]
&& -{\beta_1 \over 2 \beta_0} \Bigg(\Gqqa \Ggga-\Ggqa \Gqga\Bigg)
   +{\beta_0 \over 2} \Cq \Bigg(\Gqqa+\Ggga-2 \beta_0\Bigg)
\nonumber\\[2ex]
&&-{\beta_0 \over 2} {\Cg \over \Gqga} \Bigg[(\Gqqa)^2-\Gqqa\Ggga+2 
\Gqga\Ggqa-2 \beta_0 \Gqqa\Bigg]
\nonumber\\[2ex]
&&-{\beta_0 \over 2} \Bigg(\Gqqb-{\Gqqa \Gqgb \over \Gqga} \Bigg)
\\[2ex]
\label{eqG23}
K_{dd}^{N(1)}&=&\Gqqb+\Gggb-\bb \Bigg(\Gqqa+\Ggga \Bigg)
\nonumber\\[2ex]
&& -{2 \beta_0 \over \Gqga} \Bigg[ \Cg \Big(\Gqqa - \Ggga-2 
\beta_0\Big)-\Gqgb\Bigg]
  +4 \beta_0 \Cq -2 \beta_1~.
\end{eqnarray}
For this combination in next-to-leading order the evolution depends on 
two evolution kernels only.
In the case of polarized deep inelastic scattering similar relations 
apply considering the structure function $g_1(x,Q^2)$ and its slope. 
The anomalous dimensions and coefficient functions of the
unpolarized case have to be substituted by those for polarized scattering.

\vspace{6mm}
Although enforced by Eq.~(\ref{eqevo2}) one has still to show that
the kernels (\ref{eqG11}-\ref{eqG13}, \ref{eqG21}--\ref{eqG23}) are
scheme--independent by an explicit calculation, which we  have done
using Eqs.~(\ref{eqS3},\ref{eqS4}) for the next-to-leading order 
contributions. In leading order the scheme--invariance is visible 
explicitly, since the leading order anomalous dimensions and the lowest 
order coefficient functions for $F_L(x,Q^2)$ are scheme invariants.  

With the help of the evolution equation~(\ref{eqevo2}) we are now
prepared to ask for the validity of  crossing relations between
different space-- and time--like quantities in perturbative QCD.
Relations of this kind are henceforth called Drell--Levy--Yan (DLY)
relations,
although the original reasoning of these and other authors  
was quite different. One condition to ask such a question at all is
that the behavior of all contributing parts under crossing from
space-- to time--like momentum transfer are controlled. At large
momentum transfers $|q^2|$ the single parton picture applies and the
non--perturbative parton densities factorize. This makes it
possible to study the respective evolutions kernels without reference
to the {\sf non--perturbative}  input densities. Even if a crossing
relation for these quantities does not exist, one still may investigate
whether it exists for the {\sf perturbative} evolution kernels. A
further condition for this investigation is that the latter quantities
are {\sf scheme--invariant}, as in Eq.~(\ref{eqevo2}).
\section{\bf Drell-Levy-Yan relations}
\label{sec:DLY}

\vspace{1mm}
\noindent
In the following  we study in detail an interesting relation between deep
inelastic lepton hadron scattering and $e^+e^-$ annihilation into a 
hadron and anything else, proposed by Drell, Levy and Yan~\cite{BAS1}.  
Here we first briefly illustrate the idea behind the
work of DLY for completeness. In field theory, the deep inelastic 
$e^+e^-$ annihilation can be related to matrix elements of hadronic
electromagnetic current operators similar to that of deep inelastic 
lepton--hadron scattering. The crucial difference, apart from the ones 
which originate from the kinematics, is that the annihilation process is 
$not$ related to the forward Compton amplitude contrary to deep inelastic 
scattering because in the former process the hadron is observed in the final 
state. Nevertheless, both processes are related by crossing symmetry which any 
field theory enjoys. This motivated DLY to study the process in detail 
and then relate it to the deep inelastic scattering process. From the 
structure of the hadronic tensors $W_{\mu \nu}^S(q,p)$ (space--like) and 
$W_{\mu \nu}^T(q,p)$ (time--like) and using the standard reduction 
formalism one can infer that 
\begin{equation}
W_{\mu \nu}^T(q,p)=-W_{\mu \nu}^S(q,-p)~,
\end{equation}
where the momenta within the respective parentheses of the above 
quantities are the same as those defined in the beginning of the paper.
In the Bjorken limit for both deep inelastic scattering and deep
inelastic annihilation for $q^2 = -Q^2, p.q \rightarrow \infty$ and
$q^2 = Q^2, p.q \rightarrow \infty$, respectively
the scaling structure and fragmentation
functions satisfy the following relation~\footnote{Here we indicate
the overall signs in case of the scattering of particles of different spin,
~cf.~\cite{BUKH}. In the original work of DLY \cite{BAS1} the Yukawa-theory
was discussed which does not contain gauge bosons.}:
\begin{equation}
\label{dlyf}
F_i^{(S)}(x_B) = - (-1)^{2(s_1+s_2)} 
x_E F^{(T)}_i\left(\frac{1}{x_E}\right) \quad , \quad i=1,2,L\,.
\end{equation}
Here it has been assumed that non--perturbative input parton densities can
be decoupled trivially and are the same.
In other words, the functions  $F_i^{(T)}(x_E)$ 
are the analytic continuations
of the corresponding functions $F_i^{(S)}(x_B)$ 
from $0< x_B \leq 1$ to $1 \leq x_E < \infty$.  This is
true only when the continuation is smooth, i.e. if there are no 
singularities for example at $x=1$ etc.  This relation is called 
DLY--relation in the literature.
 
In this section, we study this property in more detail extending earlier 
work~\cite{brv}. It is particularly interesting to study the above 
transformation at the level of the
splitting functions and coefficient functions which constitute the 
physical quantities such as the structure and fragmentation functions.  
Then we show how these relations are preserved for the physical 
quantities by looking at the kernels discussed in the 
previous section. Apart from scaling violation one also encounters
distributions of the type
\begin{equation}
\delta(1-z) \qquad , \qquad \left(\frac{\ln^i(1-z)}{1-z}
\right)_+ \quad , \quad i=0,1,2, \cdots\,,
\end{equation}
which destroy the continuation  through $z = 1$. Here the distribution
$(\ln^i(1-z)/(1-z))_+$ is represented by
\begin{equation}
\left(\frac{\ln^i(1-z)}{1-z}\right)_+ =  \delta(1-z) \frac{\ln^{i+1}
\delta}{(i+1)} + \theta(1-\delta-z) \frac{\ln^i(1-z)}{(1-z)}~,
\end{equation}
where $\delta \ll 1$. It turns out that the 
DLY--relation is violated for the coefficient functions and splitting 
functions separately because both are scheme dependent. This in 
particular happens when we adopt the ${\overline {\rm MS}}$-scheme. Here 
the relation is already violated up to one-loop order for the coefficient
functions. Although one can choose other schemes in which 
Eq.~(\ref{dlyf}) is preserved (see \cite{brv}) up to one-loop order we do
not know whether this will hold up to any arbitrary order in perturbation
theory.

Let us start with the simplest examples and consider
the scheme--invariant evolution kernels Eq.~(\ref{eqG11}, \ref{eqG21}).
\subsection{\bf The Drell-Levy-Yan Relations at Leading Order}

\vspace{1mm}
\noindent
In the case of the scheme--independent evolution kernels describing the
evolution of $F_2(x,Q^2)$ and $\partial F_2(x,Q^2)/\partial t$, 
respectively, or its polarized counterpart for the structure function
$g_1(x,Q^2)$, only the transformation of two combinations of the
leading order anomalous dimensions has to be considered, 
cf.~(\ref{eqG21}). These are the determinant and the trace of the
singlet anomalous dimension matrix at leading order. In both quantities
the color factors of the off--diagonal elements enter only as a product.
The unpolarized and polarized leading order splitting functions read
\begin{eqnarray}
\label{LO2}
P_{qq}^{(0)}(z) = \Delta P_{qq}^{(0)}(z) 
&=& 4 C_F \left[\frac{1+z^2}{(1-z)_+} + \frac{3}{2}
\delta(1-z) \right]
\\[2ex]
P_{qg}^{(0)}(z) &=& 8 T_R N_f \left[z^2 + (1-z)^2\right]
\\[2ex]
\Delta P_{qg}^{(0)}(z) &=& 8 T_R N_f \left[z^2 - (1-z)^2\right]
\\[2ex]
P_{gq}^{(0)}(z) &=& 4 C_F \frac{1+(1-z)^2}{z}
\\[2ex]
\Delta P_{gq}^{(0)}(z) &=&4  C_F \frac{1-(1-z)^2}{z}
\\[2ex]
P_{gg}^{(0)}(z) &=& 8 C_A \left[\frac{z}{(1-z)_+} + \frac{1-z}{z}
+ z(1-z) \right] + 2 \beta_0 \delta(1-z) 
\\[2ex]
\label{LO3}
\Delta P_{gg}^{(0)}(z) &=& 8 C_A \left[\frac{1}{(1-z)_+} +1 -2 z \right]
+ 2 \beta_0 \delta(1-z)~.
\end{eqnarray}
The crossing relations of the leading order splitting functions are
\begin{equation}
\label{LO1}
\begin{array}{rclcrcl}
\bar P_{qq}^{(0)}&=&-z P_{qq}^{(0)}
\left (\displaystyle{1 \over z}\right)&
\hspace{1cm}& 
\bar P_{qg}^{(0)}
&=&\displaystyle{{C_F \over 2 N_f T_f}}~z P_{qg}^{(0)}
\left ({1 \over z}\right) \\ \\
\bar P_{gq}^{(0)}
&=&\displaystyle{{2 N_f T_f \over C_F}}~z P_{gq}^{(0)}
\left ({1 \over z}\right) &\hspace{1cm}&
\bar P_{gg}^{(0)}
&=&-z P_{gg}^{(0)}
\left (\displaystyle{1 \over z}\right)~,
\nonumber
\end{array}
\end{equation}
where one demands
\begin{eqnarray}
\delta(1-z) \rightarrow -\delta(1-z)~.
\label{con4}
\end{eqnarray}
Eq.~(\ref{LO1}) is easily verified 
and implies the validity of the
crossing relation from space-- to time--like evolution
kernels~Eq.~(\ref{eqG21}), i.e. the validity of the DLY--relation for
this case. For the second set of physical evolution kernels the 
DLY--relation follows at leading order referring to the transformation
relations for the leading order longitudinal coefficient functions,
Eqs.~(\ref{dlyCLq1}, \ref{dlyCLg1}) in an analogous way.
\subsection{\bf NLO Splitting function}

\vspace{1mm}
\noindent
As we know, the splitting functions and coefficient functions are not
physical quantities due to their factorization scheme dependence. Hence,
the naive continuation rule for these quantities may be violated, which 
is 
indeed the case in most of the schemes, e.g. in the $\overline{\rm MS}$ 
scheme 
characteristic of $n$-dimensional regularization.
It was demonstrated  by Curci, Furmanski and Petronzio~\cite{cfp} that
by an appropriate modification of the continuation rule in the 
$\overline{\rm MS}$ scheme one can show that the time--like splitting 
functions are related to their space--like counter parts. Since the 
modification of the continuation rule has to do with the scheme one 
adopts, it simply amounts to finding finite renormalization factors.  
It was shown that the finite renormalization factors can be constructed 
from the $\epsilon$--dependent part of the splitting function when 
computed in dimensional regularization~\cite{sv}. In addition to this,
care should be taken when dealing with quark and gluon states which was 
not the case in the work by DLY, where a color and flavor neutral field 
theory was discussed. The transformation rules are~: 
\begin{itemize}
\item
The diagonal elements of the space--like flavor singlet splitting 
functions $P_{qq}, P_{gg}$ have to be multiplied by $(-1)$.
\item
The off--diagonal elements of the singlet splitting functions matrix 
have to be multiplied by $C_F/(2 N_f T_f)$ for $P_{qg}$ and $2 N_f
T_f/C_F$ for $P_{gq}$, respectively, accounting for the interchange of 
the initial and final state particles under crossing.
\end{itemize}
 
\vspace{.1mm}
\noindent
Note that these transformations are automatically accounted for in the
case of the leading order physical evolution kernels discussed in the
previous paragraph. Keeping this in mind and using the known splitting 
functions~\cite{fp,cfp,flo} and the continuation rules
\begin{eqnarray}
&&\ln(1-z) \rightarrow \ln(1-z)-\ln(z)+i \pi\,,
\label{con1} 
\\[2ex]
&&\ln(\delta )  \rightarrow \ln(\delta) + i \pi\,,
\label{con3}
\end{eqnarray}
one finds that
\begin{eqnarray}
\label{dlysplitqq}
\bar P_{qq}^{(1)(S)}-P_{qq}^{(1)(T)}&=&
-2 \beta_0 Z^{(T)(1)}_{qq}+Z^{(T)(1)}_{qg}
\otimes \bar P_{gq}^{(0)}-Z^{(T)(1)}_{gq}
\otimes \bar P_{qg}^{(0)}\,,
\\[2ex]
\label{dlysplitqg}
\bar P_{qg}^{(1)S}-P_{gq}^{(1)(T)}&=&
-2 \beta_0 Z^{(T)(1)}_{qg}+Z^{(T)(1)}_{qg}\otimes(\bar P_{gg}^{(0)}-
\bar P_{qq}^{(0)})
+\bar P_{qg}^{(0)}\otimes(Z^{(T)(1)}_{qq}-Z^{(T)(1)}_{gg})\,,
\\[2ex]
\label{dlysplitgq}
 \bar P_{gq}^{(1)S}-P_{qg}^{(1)(T)}&=&
-2 \beta_0 Z^{(T)(1)}_{gq}+Z^{(T)(1)}_{gq}\otimes(\bar P_{qq}^{(0)}-
\bar P_{gg}^{(0)})
+\bar P_{gq}^{(0)}\otimes(Z^{(T)(1)}_{gg}-Z^{(T)(1)}_{qq})\,,
\\[2ex]
\label{dlysplitgg}
\bar P_{gg}^{(1)S}-P_{gg}^{(1)(T)}&=&
-2 \beta_0 Z^{(T)(1)}_{gg}+Z^{(T)(1)}_{gq}\otimes \bar P_{qg}^{(0)}
-Z^{(T)(1)}_{qg}\otimes \bar P_{gq}^{(0)}~,
\end{eqnarray}
where the quantities with a bar  denote that they are continued from 
$z \rightarrow 1/z$ with the appropriate factors in front. These 
quantities read in explicit form~:
\begin{equation}
\label{SPcr3}
\begin{array}{rclcrcl}
\bar P_{qq}^{(n)}(z) &=&-z P_{qq}^{(n)}\left (\displaystyle{1 \over z}\right)&
\hspace{1cm}& 
\bar P_{qg}^{(n)}(z)&=&\displaystyle{{C_F \over 2 N_f T_f}}~z P_{qg}^{(n)}
\left ({1 \over z}\right) \\ \\
\bar P_{gq}(z)^{(n)}&=&\displaystyle{{2 N_f T_f \over C_F}}~z P_{gq}^{(n)}
\left ({1 \over z}\right) &\hspace{1cm}&
\bar P_{gg}(z)^{(n)}&=&-z P_{gg}^{(n)}\left (\displaystyle{1 \over z}\right)~.
\nonumber
\end{array}
\end{equation}
The relations given in Eqs.~(\ref{dlysplitqq}--\ref{SPcr3}) 
remain true for
the polarized splitting functions~ \cite{mvv,sv} as well. The 
renormalization factors appearing in the 
Eqs.~(\ref{dlysplitqq}--\ref{dlysplitgg})
are given by
\begin{eqnarray}
\label{eqZT}
Z^{T(1)}_{ij}&=&P_{ji}^{(0)} \Big(\ln(z) + a_{ji}\Big)~.
\end{eqnarray}
The constants $a_{ij}$ are different in the unpolarized and polarized
case. For unpolarized scattering they read
\begin{equation}
\label{eqaij}
a_{qq}=a_{gg}=0\quad ,\quad a_{qg}=-\frac{1}{2}\quad , \quad 
a_{gq}=\frac{1}{2}~,
\end{equation}
whereas in the polarized case
\begin{equation}
\label{peqaij}
a_{ij}=0~.
\end{equation}
The logarithms in the renormalization factors originate from the 
kinematics. In dimensional regularization, when one continues the partonic
structure function $\hat {\cal F}_{i,k}$ in Eq. (\ref{PAR}) from the 
space--like to the time--like region one obtains an 
additional factor $z^\epsilon$ which when multiplied with the pole in 
$\epsilon$ yields $\ln(z)$. Since the pole is always associated with 
the splitting functions, one has the function $P_{ij}^{(0)}$ along with 
$\ln(z)$. The $z$--independent constant $a_{ij}$, which is also multiplied
by the splitting function, results from the
polarization average. For deep inelastic scattering one averages the
processes with one gluon in the initial state by a factor 
$1/(\epsilon+2)$. Such an average is not needed for the annihilation process 
since here the gluon appears in the final state. Notice that the average
over the polarization sum does not show up in the polarized structure functions.
Therefore in this case the constants $a_{ij}$ are zero. 

The transformation behavior of the non--singlet splitting functions
in NLO have been worked out in \cite{cfp}
where also the relations for the NLO non--singlet coefficient
functions were presented.
\subsection{\bf NLO Coefficient Functions}

\vspace{1mm}
\noindent
Now, let us study how space--like and time--like coefficient functions
are related.  The coefficient functions are expected to 
violate the DLY--relation due to their scheme dependence. Here we first 
present the relations between the space--like and time--like coefficient 
functions $C_{i,k}(z)$~($i=1,L; k=q,g$). The leading order transverse 
coefficient functions are identical. At next-to-leading order, in the 
$\overline {\rm MS}$ scheme \cite{bbdm}, the coefficient functions are 
related by the $Z$--factors in Eq. (\ref{eqZT}) as follows~:
\begin{eqnarray}
\label{dlyCTq1}
C_{1,q}^{(T)(1)}(z)+\left \{z\, C_{1,q}^{(S)(1)}\left({1 \over z}\right)
\right \}&=&Z^{(T)(1)}_{qq}
\\
\label{dlyCTg1}
{1 \over 2}\left [C_{1,g}^{(T)(1)}(z)-{C_F \over 2 N_f T_f}
\left \{2 z C_{1,g}^{(S)(1)}\left({1 \over z}\right)\right \}\right]&
=&Z^{(T)(1)}_{qg}~.
\end{eqnarray}
Since the coefficient functions depend on the hard scale of the process,
one has to replace the space--like $q^2 $ by the time--like $q^2$
in addition to Eqs.~(\ref{con1}--\ref{con3}). This leads to the
following continuation rule
\begin{eqnarray}
\label{con5}
\ln\left(\frac{Q^2}{\mu_f^2}\right)_{\rm space-like} \rightarrow
\ln\left(\frac{Q^2}{\mu_f^2}\right)_{\rm time-like} - i \pi~.
\end{eqnarray}
The $Z$--factors get contributions from two sources. The first one is
$z$--dependent and comes from the phase space integrals. The time--like 
phase space acquires an extra factor $z^\epsilon$ which gives a finite 
contribution when being multiplied with the pole terms $1/\epsilon$. The
pole term originates from the collinear divergence in $n$--dimensional
regularization. The second term originates from the polarization
average which is again absent in the time--like case. The continuation 
rules given in Eqs.~(\ref{con1}--\ref{con4}, \ref{con5}) are essential to
get the constant $\zeta(2)$ right when one goes from the space--like to the 
time--like region. Notice
that the space--like coefficient function contains
$-4 \zeta(2) \delta(1-z)$ and the time--like one contains
$8 \zeta(2) \delta(1-z)$. The difference which is $12 \zeta(2)$ can be 
understood to originate from the one-loop vertex correction
when one continues from the space--like to time--like region in $Q^2$. 
The same also holds when other regularization methods for the collinear
divergences are chosen. It is worth noticing that if one would replace
$\ln(1-z) \rightarrow \ln(1-z)-\ln(z)$ contrary to the prescription in 
Eq.~(\ref{con1}) one would obtain an additional term $12 \zeta(2)$ on the 
righthand side of Eq.~(\ref{dlyCTq1}).

The zeroth order longitudinal coefficient functions are identically zero
so that the first order contributions are scheme independent. This implies
that there are no pole terms in the corresponding partonic structure function
$\hat {\cal F}_{L,k}$. 
Hence, there is no left--over finite piece which could arise 
from the $z^\epsilon$-- or $n-$dimensional polarization average.
We find 
\begin{eqnarray}
\label{dlyCLq1}
C_{L,q}^{(T)(1)}(z)-{z \over 2}  C_{L,q}^{(S)(1)}\left({1 \over z}\right) &=& 0
\,,
\\[2ex]
\label{dlyCLg1}
{1 \over 2} \left [C_{L,g}^{(T)(1)}(z)+{C_F \over 2 N_f T_f} \left \{z \,
 C_{L,g}^{(S)(1)}\left({1 \over z}\right)\right \}\right] &=& 0~.
\end{eqnarray}
\subsection{\bf NNLO Coefficient Functions}

\vspace{1mm}
\noindent
\subsubsection{\bf Longitudinal Coefficient Functions}

\vspace{1mm}
\noindent
We consider the NNLO correction to the longitudinal coefficient
function. We follow the results given in~\cite{sanchez,zv2,zijlstra}
for the space--like and~\cite{rv2,rijken} for the time--like case. It 
turns out that the coefficient functions are related by the $Z$--factors 
through the matrix--valued convolutions
\begin{eqnarray}
\label{dlyCLq2}
\!\!\!\!C_{L,q}^{(T)(2)}(z)\!+\!\left \{-{z \over 2}\, C_{L,q}^{(S)(2)}
\left({1 \over z}\right) \right \}
\!\!\!&=&\!\!\!
\!Z^{(T)(1)}_{qq}\otimes {z \over 2} C_{Lq}^{(1)(S)}
\left({1 \over z}\right)
\nonumber\\[2ex]
&& +Z^{(T)(1)}_{gq}\otimes {C_F \over 2 N_f T_f} \left \{-{z \over 2}\,
C_{L,g}^{(S)(1)}\left({1 \over z}\right) \right \} \,,
\\[2ex]
\label{dlyCLg2}
{1 \over 2} \left[C_{L,g}^{(T)(2)}(z)\!+\!{C_F \over 2 N_f T_f} \left \{z 
 C_{L,g}^{(S)(2)}\left({1 \over z}\right)\right\}\right]
\!\!\!&=&\!\!\!
\!Z^{(T)(1)}_{qg}\otimes {z \over 2} C_{L,q}^{(S)(1)}\left({1 \over z}\right)
\nonumber\\[2ex]
&& +Z^{(T)(1)}_{gg}\otimes {C_F \over 2 N_f T_f} \left \{-{z \over 2}\,
C_{L,g}^{(1)S}\left({1 \over z}\right)\right \}~. 
\end{eqnarray}
The right hand side of the above equation contains the convolutions
of $Z$--factors with the continued NLO longitudinal space--like 
coefficient functions. We have found this pattern by comparing the scheme 
transformation which we derived in the last section. The reason for this 
structure relies on the fact that $C_{L,i}$ is obtained as the difference 
between $C_{2,i}$ and $C_{1,i}$.
Since the NLO coefficient functions involve various Nielsen--integrals,
we used the following identities to simplify the expressions~:
\begin{eqnarray}
\Li_2\left(-{1 \over z} \right)
&=&-\Li_2(-z)-{1 \over 2} \ln^2(z)-\zeta(2)\,, 
\\[2ex]
\Li_2\left(1-{1 \over z} \right)
&=&-{1 \over 2}\ln^2(z)-\Li_2(1-z)\,,
\\[2ex]
\Sf\left(1-{1 \over z} \right)
&=&-{1 \over 6}\ln^3(z)+\Sf(1-z)\,, 
\\[2ex]
\Li_3\left(1-{1 \over z} \right)
&=&{1 \over 6}\ln^3(z)+\Sf(1-z)-\Li_3(1-z)+\ln(z) \Li_2(1-z)\,,
\\[2ex]
\Li_3\left(-{1 \over z} \right)
&=&\Li_3(-z)+{1 \over 6}\ln^3(z)+\zeta(2) \ln(z)\,,
\\[2ex]
\Sf\left(-{1 \over z} \right)
&=&-\Sf(-z)+\Li_3(-z)-\ln(z) \Li_2(-z)-{1 \over 6} \ln^3(z)+\zeta(3)~.
\label{ident}
\end{eqnarray}
If we do not continue $\ln(1-z)$ and replace $\ln(1-z) \rightarrow
\ln(1-z)-\ln(z)$, then terms proportional to $\zeta(2)$ are not
compensated between space--like and time--like coefficient functions and
hence the relations given in Eqs.~(\ref{dlyCLq2}, \ref{dlyCLg2}) are no
longer true. Although formally of NNLO, the coefficient 
functions $C_{Lq(G)}^{(2)S,T}(z)$ may be combined to physical evolution
kernels together with the NLO splitting functions as shown in 
section~3.1. In Section~4.5. we will show that because of the transformation
in Eqs. (\ref{dlyCLq2}, \ref{dlyCLg2}) the physical evolution kernels in
sections 3.1 and 3.2 remain DLY--invariant.
\subsubsection{\bf Transverse Coefficient Functions}
In NNLO physical evolution kernels for the transverse structure and 
fragmentation function can only be constructed when the space--like and
time--like three-loop splitting functions are known. If they become available
one can extend Eqs.~(\ref{eqG21}--\ref{eqG23}) up to second order.
Here we consider the relation between the space-- and time--like
coefficient functions using the transformation relations (\ref{con4},
\ref{con1}, \ref{con3}) and (\ref{con5}) for unpolarized and polarized
scattering.

The space--like coefficient functions for unpolarized scattering are computed 
in \cite{zv2,zijlstra} whereas the time--like ones can be found in 
\cite{rv2,rijken}.
The transverse coefficient functions are related by (see appendix 6.1)~:
\begin{eqnarray}
C_{1,q}^{(T)(2)}(z)+ \left \{z\,
C_{1,q}^{(S)(2)}\left({1 \over z}\right)\right \} &=&
{1 \over 4}\Bigg [ 2 (-z)  P_{qq}^{(1)S}\left({1 \over z}\right) 
+ 2 \beta_0 Z^{(T)(1)}_{qq}
+  Z^{(T)(1)}_{gq} \otimes \bar P_{qg}^{(0)} 
\nonumber \\
&& - Z^{(T)(1)}_{qg} \otimes  \bar P_{gq}^{(0)} \Bigg] \ln(z)
 +{1 \over 2} Z^{(T)(1)}_{qq} \otimes  Z^{(T)(1)}_{qq} 
\nonumber \\
&& + Z^{(T)(1)}_{qq} \otimes \left(-z  C_{1,q}^{(S)(1)}\left({1 \over z}\right)
\right)
 +{1 \over 2} Z^{(T)(1)}_{gq} \otimes Z^{(T)(1)}_{qg} 
\nonumber \\
&& + Z^{(T)(1)}_{gq} \otimes \left({C_F \over 2 N_f T_f} z  C_{1,g}^{(S)(1)}
\left(\frac{1}{z}\right)\right)
+{1 \over 8} \bar P_{qg}^{(0)}\otimes \bar P_{gq}^{(0)}~.
\nonumber\\
&& +12 C_F^2 \zeta(2) \left(2 \ln\left({Q^2 \over \mu_f^2}\right)-3\right)^2
\delta(1-z)
\label{dlyCTq2}
\end{eqnarray}
For the polarized NNLO coefficient functions which were derived in 
\cite{zv2,zijlstra} and \cite{rv2,rijken}, we find that the form of
Eqs.~(\ref{dlyCTq2}) is the same but the term 
\begin{eqnarray}
{1 \over 8} \bar P_{gq}^{(0)} \otimes \bar P_{qg}^{(0)} \nonumber
\end{eqnarray}
does not occur.

Similarly for the gluonic coefficient functions we find 
\begin{eqnarray}
\label{CGNNLO}
{1 \over 2} \Bigg[C_{1,g}^{(T)(2)}(z)
-{C_F \over 2 N_f T_f} \left \{2 z\, C_{1,g}^{(S)(2)}
\left ({1 \over z}\right)\right \}\Bigg] &=&
{1 \over 4}\Bigg [ {C_F \over 2 N_f T_f}2 z  P_{qg}^{(S)(1)}
\left({1 \over z}\right) 
+ 2 \beta_0  Z^{(T)(1)}_{qg}
\nonumber\\
&& +   Z^{(T)(1)}_{qg} \otimes  \bar P_{qq}^{(0)} 
-Z^{(T)(1)}_{qq} \otimes  \bar P_{qg}^{(0)}  
 +   Z^{(T)(1)}_{gg} \otimes  \bar P_{qg}^{(0)} 
\nonumber \\
&&  -Z^{(T)(1)}_{qg} \otimes  \bar P_{gg}^{(0)} 
 \Bigg ] \Bigg(\ln(z)+{1 \over 2}\Bigg)
+{1 \over 2}  Z^{(T)(1)}_{qg} \otimes  Z^{(T)(1)}_{qq} 
\nonumber\\
&& + Z^{(T)(1)}_{qg} \otimes \left(-z  C_{1,q}^{(S)(1)}\left({1 \over z}
\right)\right)
 +{1 \over 2}  Z^{(T)(1)}_{qg} \otimes  Z^{(T)(1)}_{gg} 
\nonumber\\
&& +  Z^{(T)(1)}_{gg} \otimes \left({C_F \over 2 N_f T_f} z  C_{1,g}^{(S)(1)}
     \left({1 \over z}\right)\right)
\nonumber \\
&&-{1\over 8} \beta_0 \bar P_{qg}^{(0)}+{1 \over 16}\bar P_{qg}^{(0)} 
 \otimes \left (\bar P_{gg}^{(0)}-\bar P_{qq}^{(0)}\right)~.
 \label{dlyCTg2}
\end{eqnarray}
For the polarized case, the terms
\begin{eqnarray}
-{1\over 8} \beta_0 \bar P_{qg}^{(0)}+{1 \over 16}\bar P_{qg}^{(0)}
\otimes \left (\bar P_{gg}^{(0)}-\bar P_{qq}^{(0)}\right) \nonumber
\end{eqnarray}
in Eq.~(\ref{dlyCTg2}) are absent.
Since we do not have to average over the initial state gluon polarization
in the case of polarized scattering the term
$\ln(z)+1/2$ multiplying the first bracket in Eq.~(\ref{CGNNLO})
is replaced by $\ln(z)$, cf. also Eq.~(\ref{eqZT}, \ref{peqaij}).

\subsection{\bf NLO Physical Evolution Kernels}

\vspace{1mm}
\noindent
After having found the relations between space--like and time--like
splitting  and coefficient functions, 
we investigate the DLY-transformation
for the physical evolution kernels
presented in sections
3.1 and 3.2. In order to do this, we define the difference 
between the time--like quantities $K_{IJ}^T$ and the continued 
space--like quantities $\bar{K}_{ij}^S$ by
\begin{equation}
\delta K_{IJ}=K_{IJ}^T - \bar K_{IJ}^S~,
\end{equation}
where $\bar K_{IJ}^S$ is obtained by transforming $K_{IJ}^S$ to the time--like 
region using the continuation rules~(\ref{con1}, \ref{con3}, \ref{con4}, 
\ref{con5}).
Application of the DLY--transformations provides us with the following results
\begin{eqnarray}
\delta K_{22}^{N(1)}&=&\delta \Gqqb - \bCLra \delta \Ggqb + \bCLra 
\delta \Cg \bar \Gqqa
\nonumber\\[2ex]
&& - \left[{\delta \CLqb \over \bar \CLga}+\left(\bCLra\right)^2 \delta \Cg 
   -\bCLra {\delta \CLgb \over \bar \CLga} \right] \bar \Gqga
\nonumber\\[2ex]
&& +\bar \Ggqa \delta \Cg-\bCLra \bar \Ggga \delta \Cg 
   + 2 \beta_0 \left(\delta \Cq -\bCLra \delta \Cg \right) 
\nonumber\\[2ex]
 &=& \delta \Gqqb - 2 \beta_0 \ZqqNT - \bar \Ggqa  \ZqgNT+\bar \Gqga \ZgqNT
\nonumber\\[2ex]
&&+\bCLra ( -\delta \Ggqb +2 \beta_0 \ZqgNT -\ZqgNT \bar \Gqqa + \ZqqNT 
\bar \Gqga 
\nonumber\\[2ex]
&&  -\ZggNT \bar \Gqga +\ZqgNT \bar \Ggga ) \,.
\label{delgam22}
\end{eqnarray}
Substituting the expressions for $\delta \Gqqb$ and $\delta \Ggqb$ using
Eqs.~(\ref{dlysplitqq}, \ref{dlysplitgq}), we get
\begin{equation}
 \delta K_{22}^{N(1)} = 0~.
\end{equation}
For the remaining evolution kernels one obtains
\begin{eqnarray}
 \delta K_{2L}^{N(1)}&=& \delta \Ggqb -2 \beta_0 \ZqgNT +\ZqgNT 
(\bar \Gqqa-\bar \Ggga)
                        -\bar \Gqga (\ZqqNT-\ZggNT)\,,
\label{delgam2L}
\\[2ex]
 \delta K_{LL}^{N(1)}&=&\delta \Gggb -2 \beta_0 \bar \Ggga 
                          +\ZqgNT \bar \Ggqa - \ZgqNT \bar \Gqga
\nonumber\\[2ex]
&& +\bCLra \left [ \delta \Ggqb-2 \beta_0 \ZqgNT + \ZqgNT \bar \Gqqa
   -\ZqgNT \bar \Ggga - \ZqqNT \bar \Gqga \right.
\nonumber\\[2ex]
&& \left. + \ZggNT \bar \Gqga \right]\,,
\label{delgamLL}
\\[2ex]
\delta K_{L2}^{N(1)}&=& \delta \Gqgb+\bCLra \delta \Gqqb 
-\left(\bCLra\right)^2 \delta \Ggqb
                          -\bCLra \delta \Gggb
\nonumber\\[2ex]
&& -2 \beta_0 \ZgqNT-\ZgqNT \bar \Gqqa + \ZqqNT \bar \Ggqa -\ZggNT 
\bar \Ggqa +\ZgqNT \bar \Ggga
\nonumber\\[2ex]
&&+ \bCLra \left[ -2 \beta_0 \ZqqNT -\ZqgNT \bar \Ggqa + \ZgqNT 
\bar \Gqga \right]
\nonumber\\[2ex]
&&+\left ( \bCLra \right)^2 \left[ 2 \beta_0 \ZqgNT -\ZqgNT \bar 
\Gqqa -\ZggNT \bar \Gqga +\ZqqNT \bar \Gqga \right.
\nonumber\\[2ex]
&& \left. +\ZqgNT \bar \Ggga \right]
+ \bCLra \left[ 2 \beta_0 \ZggNT -\ZqgNT \bar \Ggqa +\ZgqNT \bar 
\Gqga \right]~.
\label{delgamL2}
\end{eqnarray}
The explicit expressions for the differences in the coefficient function
are given in appendix 6.1 as well as a series of involved 
Mellin--convolutions leading to Nielsen--integrals (see appendix 6.2), 
which are necessary in the explicit calculation.

Using Eqs.~(\ref{dlysplitqq}--\ref{dlysplitgg}), leads to
\begin{eqnarray}
 \delta K_{L2}^{N(1)} &=& 0 \,, 
\\[2ex]
 \delta K_{2L}^{N(1)} &=& 0 \,, 
\\[2ex]
 \delta K_{LL}^{N(1)} &=& 0~.
\end{eqnarray}
The physical evolution kernels    $K_{I,J}$ for the evolution
of the structure functions $F_2$ and $F_L$ are thus DLY--invariant to
next-to-leading order if continued from the space--like to the 
time--like region.

We turn now to the physical evolution kernels    in next-to-leading
order where we choose the physical quantities
$F_2$, $\partial F_2 / \partial t$ as a basis. Here only
two evolution kernels are contributing, which change under the 
DLY-transformation as follows~:
\begin{eqnarray}
\delta K_{d2}&=& {\beta_0 \over 2}\Bigg(\delta C_{2q}^{N(1)} 
-\ZqqNT\Bigg)
\bigg(\bar \Gqqa+\bar \Ggga-2 \beta_0\bigg)
\nonumber\\[2ex]
&&-{\beta_0 \over 2 \bar \Ggqa}\Bigg(\delta C_{2g}^{N(1)} 
-\ZqgNT\Bigg) \Bigg((\bar \Gqqa)^2-\bar \Gqqa \bar \Ggga
\nonumber\\[2ex]
&&+2 \bar \Gqga \bar \Ggqa -2 \beta_0 \bar \Gqqa\Bigg)
\label{delgamd2}
\\[2ex]
\delta K_{dd}&=&-~2 { \beta_0 \over \bar \Ggqa} 
\Bigg(\delta C_{2g}^{N(1)}-\ZqgNT\Bigg)
\bigg(\bar \Gqqa-\bar \Ggga-2 \beta_0\bigg)
\nonumber\\[2ex]
&&+4 \beta_0 \bigg(\delta C_{1q}^{N(1)}-\ZqqNT\bigg)\,.
\label{delgamdd}
\end{eqnarray}
From Eqs. ~(\ref{dlyCTq1}, \ref{dlyCTg1}, \ref{dlyCLq1}, \ref{dlyCLg1}) we
can derive that
\begin{eqnarray}
\delta K_{d2} &=& 0 \,,
\\[2ex]
\delta K_{dd} &=& 0~.
\end{eqnarray}
From these results it is clear
that the time--like physical evolution kernels $K_{ij}^T$ can be directly 
derived from the space--like physical evolution kernels using the
continuations in Eqs.~(\ref{con1}, \ref{con3}, \ref{con4}, \ref{con5})
where one has to account for the corresponding changes in the overall color 
factors. The $Z^T$--factors which are needed for the transformation of the
splitting and coefficient functions cancel in the expression above.
In the future one can extend the investigation performed in this section
to physical evolution kernels at the NNLO-level, provided the 3--loop
anomalous dimensions are calculated. For the choice of observables 
$(F_2,F_L)$ one also needs the three-loop coefficient functions.

\vspace{2mm}
We finally would like to comment on a relation derived by
Gribov and Lipatov in \cite{GL} for the leading
order kernels for a pseudoscalar and a vector field theory.~\footnote{
See also \cite{REL} for related work.}
One may write it in the form
\begin{equation}
\overline{K}(x_E,Q^2) = K(x_B,Q^2)~,
\end{equation}
where $\overline{K}$ and $K$ denote the time-- and space--like
evolution kernels, respectively, and $x_B = 1/x_E$. One
verifies, that this relation holds in leading order for the space--
and time--like splitting functions of QCD, Eqs.~(\ref{LO2}--\ref{LO3}),
without changing the $\delta$--function, Eq.~(\ref{con4}).

Starting with next-to-leading order, this relation is not preserved. For
the physical non--singlet evolution kernels this was shown in 
\cite{cfp,flo} and for some singlet combinations in~\cite{flo}.
We find, that also for the physical singlet
combinations, Eqs.~(\ref{eqG12}--\ref{eqG13}, \ref{eqG22}--\ref{eqG23}),
this relation is violated as well.
\section{\bf Conclusions}

\vspace{1mm}
\noindent
The old question, whether the scattering cross sections of deep inelastic 
scattering $e^- + P \rightarrow e^- + `X'$ are related to the annihilation 
cross section $e^+ + e^- \rightarrow \bar P + `X'$ by a crossing relation
changing from $t$-- to $s$--channel was newly discussed. Since in both 
reactions non--perturbative quantities such as the structure and 
fragmentation functions contribute the above question cannot be answered 
by means of perturbation theory for the process as a whole. However, since 
both the parton densities involved in the space--  and time--like process
factorize if the virtuality $Q^2 = |q^2|$ of the four--momentum transfer 
is large a related question can be asked for the crossing behavior of the 
respective evolution kernels, which are computable within perturbation 
theory. In the calculation of both inclusive processes only two types of 
singularities occur, the collinear singularity and the ultraviolet 
singularity. These divergences are absorbed into the bare parton densities 
and the coupling constant, respectively. Two distinct renormalization 
group equations are implied. They quantify the impact of the factorization 
and the renormalization scale on the DIS structure functions and 
fragmentation functions when the perturbation series is truncated up to a
given order. However one can construct factorization--scale independent 
evolution kernels which describe the scheme--invariant evolution of these 
physical quantities in terms of a kinematic   variable given by $Q^2$. 
This scheme invariant evolution is guaranteed up to any finite order in 
perturbation theory. Notice that in finite order this method does not 
remove the dependence of the physical quantities on the renormalization 
scheme of the strong coupling constant or its scale $\mu_r$.

The first example of the application of the physical evolution kernels is the 
coupled structure functions $F_2(x,Q^2)$ and $F_L(x,Q^2)$ associated with
the corresponding fragmentation functions in $e^+e^-$--annihilation. A 
second example is given by $F_2(x,Q^2)$ and $\partial F_2(x,Q^2)/\partial
\ln(Q^2)$. Contrary to the splitting functions (anomalous dimensions) and
coefficient functions the evolution kernels of the examples above are
factorization scheme independent. For that purpose transformation 
relations have been derived for the splitting functions up to NLO and the 
coefficient functions up to NNLO. We have also shown that these kernels 
are invariant under the Drell--Levy--Yan--transformation up to 
next--to--leading order. On the other hand the Gribov--Lipatov relation, 
which is valid in leading order, is already violated at next-to-leading 
order. It remains to be seen how the physical evolution kernels behave 
under the DLY crossing relation at NNLO, which presumes the knowledge of 
the yet unknown three-loop splitting functions (space-- and time--like) 
as well the three--loop longitudinal coefficient functions in the first 
example above.

\vspace{2mm}
\noindent
{\bf Acknowledgment.}~~We would like to thank P.~Menotti for providing
us a reprint of \cite{Men}. Discussions with S. Kurth in an early phase
of this work are acknowledged. This work was supported in part by EU 
contract FMRX-CT98-0194 (DG 12-MIHT).
\section{Appendix}
\subsection{Coefficient Functions}

\vspace{1mm}
\noindent
In this appendix we list the difference of the space- and time--like
coefficient functions in the $\overline{\rm MS}$ scheme, which are used
in section~\ref{sec:DLY} to study the validity of the DLY--relation. Here the
expressions also contain the logarithms
\begin{equation}
\label{LOG}
L_{\mu_f} = \ln(Q^2/\mu_f^2)
\end{equation}
which arise when the factorization scale $\mu_f^2$ is chosen to be different
from $Q^2$.

The difference between the longitudinal non--singlet 
coefficient functions
corresponding to the processes $\gamma^* + q \rightarrow q + g + g$
and $\gamma^* \rightarrow `\bar q' + q + g + g$ respectively are given by
\begin{eqnarray}
\label{CLqNS}
\delta C_{L,q}^{(2)NS}&=& C_F^2~ \Big [~4 \Big(2 z -\ln(z)\Big)
\ln(z)
-16 \Li_2(1-z)
+8-8 z \Big]
\end{eqnarray}
where $`q'$ denotes the quark in the final state which undergoes fragmentation
into a hadron $P$ (see Eq. (8)).
The difference between the longitudinal purely singlet coefficient functions
corresponding to the processes $\gamma^* + q \rightarrow q + q + \bar q$
and $\gamma^* \rightarrow `\bar q' + q + q + \bar q$ respectively are given by
\begin{eqnarray}
\label{CLqPS}
\delta C_{L,q}^{(2)PS}&=& 
N_f T_f C_F \left[ -8 \left(6 +4 z -{4 \over 3} z^2\right) \ln(z)
+16 \ln^2(z) -16 
\right.  
\nonumber\\[2ex] 
&&\left. -{304 \over 9 z} +64 z-{128 \over 9} z^2 \right]
\end{eqnarray}
The same is done for
the longitudinal gluonic coefficient functions
corresponding to the processes $\gamma^* + g \rightarrow g + q + \bar q$
and $\gamma^* \rightarrow `g' + g + q + \bar q$, respectively. The 
difference in the coefficient function is
     given by
\begin{eqnarray}
\label{CLg}
\delta C_{L,g}^{(2)}&=& C_F^2 \left[8 \left(1 +{2 \over z} -z 
\right) \ln(z)
+8 \ln^2(z)-28 +{24 \over z} + 4 z \right]
\nonumber\\[2ex]
&& +C_A C_F \left[ 16 \left(4 - {2 \over z} + z -{1 \over 3} z^2 
\right) \ln(z)
-16 \left(1 + {1 \over z}\right) \ln^2(z) \right.
\nonumber\\[2ex]
&& \left.
+ 32 \left(1-{1 \over z}\right) \Li_2(1-z)
-8+{248 \over 9 z} -24 z +{40 \over 9} z^2 
\right]~.
\end{eqnarray}
Notice that for the computation of the coefficients functions above
and the ones following hereafter one also needs the virtual contributions
to the zeroth and first order partonic processes.
The differences between the transverse coefficient functions emerge from the 
same processes as mentioned above Eqs. (\ref{CLqNS}, \ref{CLqPS},
\ref{CLg}). In the non--singlet case we have
\begin{eqnarray}
\label{C1qNS}
\delta C_{1,q}^{(2)NS} & = & \left( C_F^2 - {1 \over 2} C_A C_F 
\right)\left[
8{\ln(z) \over 1+z} \Bigg(-2 \zeta(2) -4 \ln(z) \ln(1+z)
+\ln^2(z) \Bigg)
\right . 
\nonumber\\[2ex]
&& +\left. 4 \Bigg( 2 (1-z)\zeta(2) +4 (1-z) +4 (1-z) \ln(z) 
\ln(1+z)
\right. 
\nonumber\\[2ex]
&&\left. +2 (1+z) \ln(z)-(1-z) \ln^2(z) \Bigg) \ln(z) -16 \Li_2(-z) \ln(z)
\right.
\nonumber\\[2ex]
&&\left.  \times \left ( {2 \over 1+z}-1+z\right) \right]
\nonumber  \\
&& 
+ N_f T_f C_F \left[ {8 \over 9} \left(-{10 \over 1-z}-1 
+11 z\right) \ln(z)
\right]
\nonumber\\[2ex]
&& +C_A C_F \left[4 {\ln(z) \over 1-z} \left({67 \over 9 } 
+ \ln^2(z) -2 \zeta(2) \right)
+2 \Bigg({53 \over 9} -{187 \over 9} z \right. 
\nonumber\\[2ex]
&&\left.  + 2 (1+z) \ln(z) - (1+z) \ln^2(z)\Bigg) \ln(z)
+4 \zeta(2) (1+z) \ln(z) 
\right] 
\nonumber\\[2ex]
&& +C_F^2 \left[4 { \ln(z) \over 1-z}  \Bigg( \Big(8 L_{\mu_f} -6 
+4 \ln(1-z) \Big)\ln(1-z) +6 L_{\mu_f} -18 \right. 
\nonumber \\
&&\left. + \Big(-4 L_{\mu_f} +6 - {16 \over 3} \ln(z) \Big ) \ln(z)\Bigg)
+4 \Bigg(-4 (1+z) L_{\mu_f}+ 1+5 z \right.
\nonumber\\[2ex]
&&\left.
-2 (1+z) \ln(1-z) \Bigg)\ln(z) \ln(1-z ) 
+2 \Bigg(-2(5+z) L_{\mu_f} +14+40 z \right.
\nonumber\\[2ex]
&& \left.  +2 (1+z) \ln(z) \ln(1-z) +6 (1+z) L_{\mu_f} \ln(z)-8 (2+z)\ln(z)
\right.
\nonumber\\[2ex]
&&\left.  +7 (1+z) \ln^2(z) \Bigg) \ln(z)
+2 \Li_2(1-z) \Bigg( 4 \left(-{12 \over 1-z}+7 + 7 z\right)\ln(z) \right. 
\nonumber\\[2ex]
&&\left. -10+22 z\Bigg)
+8 \left(-{24 \over 1-z} +13 +13 z\right)\Sf(1-z) +36 (1-z) \right.
\nonumber\\[2ex]
&& \left.
+32 \zeta(2) \left({2 \over 1-z}-1-z\right) \ln(z)
+12 \zeta(2) (9-12 L_{\mu_f}+4 L^2_{\mu_f})\delta(1-z) \right]~.
\end{eqnarray}
For the purely singlet difference we obtain
\begin{eqnarray}
\label{C1qPS}
\delta C_{1,q}^{(2)PS} & = &  N_f T_f C_F \left[ \left( -8 \Big(4+6 z 
+ {8 \over 3} z^2\Big) \Big(L_{\mu_f}+\ln(1-z)\Big) -16 \left(1-{2 \over 3 z}
 +2 z\right) \right.\right.
\nonumber\\[2ex]
&&\left.\left. \times \ln(z)-160 -{160 \over 9 z} 
-112 z -{368 \over 9} z^2+4 (1+z) \bigg(4 \ln(1-z)+4 L_{\mu_f} \right.\right.  
\nonumber\\[2ex]
&& \left.\left. +{10 \over 3} 
\ln(z)\bigg)\ln(z) \right)\ln(z)-8 \left(1+{38 \over 9 z} -z 
-{38 \over 9 } z^2 \right) \Bigg(L_{\mu_f} + \ln(1-z)\Bigg) \right.
\nonumber\\[2ex]
&&\left. +16 \left( 2(1+z) \ln(z)-2-3 z -{4 \over 3} z^2 \right)\Li_2(1-z)
+32 (1+z) \Sf(1-z) \right.
\nonumber\\[2ex]
&&\left.  - {1168 \over 9} -{224 \over 27 z} +{640 \over 9 } z
+{1808 \over 27} z^2 \right]~.
\end{eqnarray}
The difference between the gluonic coefficient functions equals
\begin{eqnarray}
\label{C1g}
\delta C_{1,g}^{(2)}&=&2 C_A C_F\left[ \left( 188 + {704 \over 9 z}
+66 z +{184 \over 9 } z^2 +\left(48 -{24 \over z} +16 z +{32 \over 3 }
z^2\right) L_{\mu_f} \right.  \right.
\nonumber\\[2ex]
&&\left.\left. +\left(-4-{100 \over 3 z} +10 z\right) \ln(z)
+\left(40-{16 \over z} +12 z +{32 \over 3 } z^2 \right) \ln(1-z)\right.\right.
\nonumber\\[2ex]
&& \left.\left.  +4 \left(2 +{2 \over z} +z \right) \ln(1+z)
-16 \left(1+{1 \over z} +z \right) \ln(z) L_{\mu_f} \right.\right.
\nonumber\\[2ex]
&&\left.\left.  +8 \left( 2 +{2 \over z} +z \right) \ln(z) \ln(1+z)
-16 \left(1+ {1 \over z} +z \right) \ln(z) \ln(1-z) \right.\right.
\nonumber\\[2ex]
&&\left.\left.  -\left( {40 \over 3} +{64 \over 3 z} + {52 \over 3 }
z\right) \ln^2(z) +4 \left(2 -{2 \over z} -z\right) \ln^2(1-z) \right)
\ln(z) \right.
\nonumber\\[2ex]
&&\left.  +\left( -32 +{356 \over 9 z} + 4 z -{104 \over 9} z^2\right) \ln(1-z)
+2 \left(2 - {2 \over z} -z\right) \ln^2(1-z) \right.
\nonumber\\[2ex]
&&\left.  +\left(-24 +{284 \over 9 z} +4 z -{104 \over 9} z^2\right) L_{\mu_f}
+\left( 32 \left(1-{3\over z}-2 z\right) \ln(z)
\right.\right.
\nonumber\\[2ex]
&&\left. \left.  +16 \left(2-{2 \over z}-z\right)\bigg(\ln(1-z)+L_{\mu_f}\bigg)
+16+{8 \over z}+16 z+{32 \over 3} z^2 \right) \Li_2(1-z) \right.
\nonumber\\[2ex]
&&\left.  +4 \left(2 + {2 \over z} +z \right)\bigg(1+2 \ln(z)\bigg)\Li_2(-z)
+16 \left(-2+{2 \over z} +z \right) \Li_3(1-z) \right.
\nonumber\\[2ex]
&&\left.  +8 \left(8-{16 \over z} - 10 z\right)\Sf(1-z)
-{418 \over 9} +{4168 \over 27 z} -{668 \over 9 } z-{856 \over 27} z^2 \right.
\nonumber\\[2ex]
&&\left.  +8 \zeta(2) \left(-3+{4 \over z} +2 z\right) (1+2 \ln(z)) \right]
\nonumber\\[2ex]
&& +2 C_F^2 \left[\left(42 +{8 \over z} -29 z-(14-11 z) \ln(z)
+16 \left(3 -{3 \over z}-z\right) \ln(1-z) \right.\right.
\nonumber\\[2ex]
&&\left.\left.  +4 (2-z)\ln(z) L_{\mu_f} -4 \left(2 -{4 \over z}-z\right)
\ln(z) \ln(1-z) \right.\right.
\nonumber\\[2ex]
&&\left.\left.  -16 \left(2 -{2 \over z} -z\right) \ln(1-z) L_{\mu_f}
+{10 \over 3} (2-z)\ln^2(z) \right.\right.
\nonumber\\[2ex]
&&\left. \left.  -12 \left(2 -{2 \over z}-z\right) \ln^2(1-z) \right)\ln(z)
+\left(64 -{52 \over z} -18 z \right.\right.
\nonumber\\[2ex]
&&\left.\left.  -8 \left(2 -{2 \over z}-z \right) L_{\mu_f}
-6 \left(2 -{2\over z}-z\right) \ln(1-z)\right)\ln(1-z) \right.
\nonumber\\[2ex]
&&\left.  +2 \left(16-{10 \over z} -3 z\right) L_{\mu_f}
+8  \left((2-z) \ln(z) -2\left( 2 - {2 \over z} -z\right)\bigg(\ln(1-z)
\right.\right.
\nonumber\\[2ex]
&&\left.\left.  +L_{\mu_f}\bigg) +6-{6 \over z}-3 z\right) \Li_2(1-z)
+16  \left(2- {2 \over z} -z\right) \Li_3(1-z) \right.
\nonumber\\[2ex]
&&\left.  +8 \left(10-{8\over z} -5 z\right)\Sf(1-z)
-169 +{106 \over z} +50 z \right]~.
\end{eqnarray}
We have computed the same differences between the coefficient functions
corresponding to the structure function $g_1(x,Q^2)$ which describes
polarized scattering. The analogues of Eqs.~(\ref{C1qNS},  \ref{C1qPS},
 \ref{C1g}) are given by
\begin{eqnarray}
\label{DC1qNS}
\delta \Delta C_{1,q}^{(2)NS}&=& \left( C_F^2 - {1 \over 2} C_A C_F 
\right)\left[ 8{\ln(z) \over 1+z} \Bigg(-2 \zeta(2) -4 \ln(z) \ln(1+z)
+\ln^2(z) \Bigg) \right . 
\nonumber\\[2ex]
&& +\left. 4 \Bigg( 2 (1-z)\zeta(2) +4 (1-z) +4 (1-z) \ln(z) 
\ln(1+z) \right. 
\nonumber\\[2ex]
&&\left. +2 (1+z) \ln(z)-(1-z) \ln^2(z) \Bigg) \ln(z)
-16 \Li_2(-z) \ln(z) \right.
\nonumber\\[2ex]
&&\left.  \times \left ( {2 \over 1+z}-1+z\right) \right]
\nonumber\\[2ex]
&& + N_f T_f C_F \left[ {8 \over 9} \left(-{10 \over 1-z}-1 +11 z\right) 
\ln(z) \right ]
\nonumber\\[2ex]
&& +C_A C_F \left [4 {\ln(z) \over 1-z} \left({67 \over 9 } 
+ \ln^2(z) -2 \zeta(2) \right) +2 \Bigg({53 \over 9} -{187 \over 9} z \right. 
\nonumber\\[2ex]
&&\left.  + 2 (1+z) \ln(z) - (1+z) \ln^2(z)\Bigg) \ln(z)
+4 \zeta(2) (1+z) \ln(z) \right ] 
\nonumber\\[2ex]
&& +C_F^2 \left[4 { \ln(z) \over 1-z}  \Bigg( \Big(8 L_{\mu_f} -6 
+4 \ln(1-z) \Big)\ln(1-z) +6 L_{\mu_f} -18 \right. 
\nonumber\\[2ex]
&&\left. + \Big(-4 L_{\mu_f} +6 - {16 \over 3} \lg(z) \Big ) \ln(z)\Bigg)
+4 \Bigg(-4 (1+z) L_{\mu_f}+ 1+5 z \right.
\nonumber\\[2ex]
&&\left.  -2 (1+z) \ln(1-z) \Bigg)\ln(z) \ln(1-z ) 
+2 \Bigg(-2(5+z) L_{\mu_f} +18+36 z \right.
\nonumber\\[2ex]
&& \left.  +2 (1+z) \ln(z) \ln(1-z) +6 (1+z) L_{\mu_f} \ln(z)-2 (7+5 z)
 \ln(z) \right.
\nonumber\\[2ex]
&&\left.  +7 (1+z) \ln^2(z) \Bigg) \ln(z)
+2 \Li_2(1-z) \Bigg( 4 \left(-{12 \over 1-z}+7 + 7 z\right)\ln(z) \right. 
\nonumber\\[2ex]
&&\left. -2+14 z\Bigg) +8 \left(-{24 \over 1-z} +13 +13 z\right)\Sf(1-z) 
+36 (1-z) \right.
\nonumber\\[2ex]
&& \left.  +32 \zeta(2) \left({2 \over 1-z}-1-z\right) \ln(z)
+12 \zeta(2) (9-12 L_{\mu_f}+4 L_{\mu_f}^2)\delta(1-z) \right]
\end{eqnarray}
\begin{eqnarray}
\label{DC1qPS}
\delta \Delta C_{1,q}^{(2)PS}&=& N_f T_f C_F \Bigg[  
\Bigg( 16 (1-4 z)\bigg(L_{\mu_f}+\ln(1-z) \bigg)-184+24 z -16 (2+3 z) 
\nonumber\\[2ex]
&&  \times \ln(z) + 4 (1+z)  \left(4 L_{\mu_f} +4 \ln(1-z)
+{10 \over 3 }\ln(z)\right)\ln(z) \Bigg) \ln(z)
\nonumber\\[2ex]
&& -48 (1-z) \bigg(L_{\mu_f} + \ln(1-z) \bigg)
+16 \bigg(2 (1+z) \ln(z)+1-4 z\bigg) \Li_2(1-z)
\nonumber\\[2ex]
&& +32(1+z) \Sf(1-z) -112+112 z \Bigg]
\end{eqnarray}
\begin{eqnarray}
\label{DC1g}
\delta \Delta C_{1,g}^{(2)}&=&2 C_A C_F \Bigg[  \Bigg(212 +48 z -32
(1-2 z) L_{\mu_f} +4 (8+5 z) \ln(z) -16 (1- 3 z) 
\nonumber\\[2ex]
&& \times \ln(1-z) -8 (4+z) \ln(z) L_{\mu_f} +8 (2 +z ) \ln(z) \ln(1+z) 
\nonumber\\[2ex]
&& -8 (4+z) \ln(z) \ln(1-z) -{4 \over 3} (26 + 5 z) \ln^2(z)-4 (2-z) 
\nonumber\\[2ex]
&& \times \ln^2(1-z) \Bigg) \ln(z) +32 (1-z) \bigg(\ln(1-z)+L_{\mu_f}\bigg) 
\nonumber\\[2ex]
&& +16 \Bigg(2+z-(2-z)\bigg(L_{\mu_f} +\ln(1-z)\bigg) -(8-z) \ln(z) 
\Bigg)\Li_2(1-z)
\nonumber\\[2ex]
&& +8 (2+z)  \ln(z) \Li_2(-z) +16 (2-z) \Li_3(1-z)-32 (5-z) \Sf(1-z)
\nonumber\\[2ex]
&& +224 -224 z +8\zeta(2) (8-3 z) \ln(z) \Bigg]
\nonumber\\[2ex]
&& +2 C_F^2 \Bigg[\bigg( 22 + 40 z +4 (2-z) L_{\mu_f}-4 (14-9 z) \ln(1-z)
+4 (8-5 z) \ln(z)
\nonumber\\[2ex]
&& +16(2-z) \ln(1-z) L_{\mu_f}-4 (2-z)  \ln(z) L_{\mu_f}
\nonumber\\[2ex]
&& +4 (2-z) \ln(z) \ln(1-z) -{10 \over 3} (2-z) \ln^2(z) 
\nonumber\\[2ex]
&& +12(2-z) \ln^2(1-z) \bigg) \ln(z) -8 (1-z) \bigg(\ln(1-z)+L_{\mu_f}\bigg) 
\nonumber\\[2ex]
&& +4 \Bigg(-10+5 z+4 (2-z)\bigg(\ln(1-z)+L_{\mu_f}\bigg)
-2(2-z) \ln(z) \Bigg)
\nonumber\\[2ex]
&& 
\times \Li_2(1-z) -16 (2-z)\Li_3(1-z)-40 (2-z) \Sf(1-z)+96-96 z \Bigg]~.
\end{eqnarray}
\subsection{\bf Convolutions}
Here we list the convolutions of a series of functions, which are
needed for the investigation of the DLY--relation in 
section~\ref{sec:DLY}. Using the definition in Eq. (\ref{CONV}) we obtain
\begin{eqnarray}
{\ln(z) \over (1-z)}\otimes {\ln(z)  \over (1-z) }
&=&
{1 \over (1-z)} \Bigg [-4~ {\rm S}_{1,2}(1-z)
                                   -2 \ln(z) {\rm Li}_2(1-z)
\nonumber \\ &&                                   
-{1  \over 6} \ln^3(z) \Bigg]
          \\
{\ln(z) \over (1-z)}\otimes {\ln(z)  \over z }&=&-~{1 \over z} 
\left[2~ {\rm S}_{1,2}(1-z)
                                         +\ln(z) {\rm Li}_2(1-z)
\right]
          \\
{\ln(z) \over (1-z)}\otimes z^2 \ln(z) &=&{1 \over 12}~ \Bigg 
[-3-24 z+27 z^2-24~ {\rm S}_{1,2}(1-z) z^2
\nonumber \\
&&-3 (1+4 z+5 z^2) \ln(z) -2 z^2 \ln^3(z)
\nonumber \\
&&-12 z^2 \ln(z) {\rm Li}_2(1-z) \Bigg]
          \\
{\ln(z) \over (1-z)}\otimes z \ln(z) &=&-2+2 z-2 z~ {\rm S}_{1,2}
(1-z)-\ln(z)-z \ln(z)
 \nonumber \\ && 
-{z \over 6} \ln^3(z)
-z \ln(z) {\rm Li}_2(1-z)  
          \\
{\ln(z) \over (1-z)}\otimes \ln(z) &=&-2~ {\rm S}_{1,2}(1-z)
-{1 \over 6} \ln^3(z)-\ln(z) {\rm Li}_2(1-z)  
          \\
{\ln(z) \over (1-z)}\otimes {\ln(1-z)  \over z} &=&
{1  
\over z} \left[-~{\rm S}_{1,2}(1-z)
                          -\ln(1-z) {\rm Li}_2(1-z)+
{\rm Li}_3(1-z)\right]
 \\
{\ln(z) \over (1-z)}\otimes z^2 \ln(1-z) &=&
{1  \over 4} 
\Bigg\{-4-5 z+9 z^2-8~ {\rm S}_{1,2}(1-z) z^2
                                  +\ln(1-z)
\nonumber \\&& 
+4 z \ln(1-z) -5 z^2 \ln(1-z)-3 \ln(z)-4 z \ln(z)
\nonumber \\&& 
+2 \ln(1-z) \ln(z) +4 z \ln(1-z) \ln(z)
\nonumber \\&& 
-2 z^2 \ln(1-z) \ln^2(z) +\bigg[2+4 z-4 z^2 \ln(1-z)
\nonumber \\&& 
-4 z^2 \ln(z)\bigg] {\rm Li}_2(1-z)
                              +4 z^2 {\rm Li}_3(1-z) \Bigg\}
 \\
{\ln(z) \over (1-z)}\otimes z \ln(1-z) &=&
-2+2 z-2 z~ 
{\rm S}_{1,2}(1-z)+\ln(1-z)-z \ln(1-z) 
\nonumber \\ &&
-\ln(z) +\ln(1-z) \ln(z)-{z  \over 2} \ln(1-z) \ln^2(z)
\nonumber \\ &&
                                -\Bigg[-1+z \ln(1-z)
+z \ln(z)\Bigg] {\rm Li}_2(1-z)
\nonumber \\&&
+z {\rm Li}_3(1-z)  
\\
{\ln(z) \over (1-z)}\otimes \ln(1-z) &=&
-2~ 
{\rm S}_{1,2}(1-z)-{1  \over 2} \ln(1-z) \ln^2(z)
\nonumber \\ && 
                      - \left[\ln(1-z)+\ln(z)\right]
{\rm Li}_2(1-z) +{\rm Li}_3(1-z)  
          \\
\!\!\!{\ln(z) \over (1-z)}
\otimes \left({\ln(1-z)\over 1-z }\right)_+ \!\!\!\!
&=&\!\!\!\!{1 \over (1-z)}
                                        \left [{1  \over 2}
\ln(z) \ln^2(1-z)
                                          -2~ {\rm S}_{1,2}(1-z)
\right. \nonumber \\ && \left.
                             -\ln(z) {\rm Li}_2(1-z) -{1  \over 2} 
                             \ln^2(z) \ln(1-z)\right]
\\
{\ln(z) \over (1-z)}
\otimes {1  \over (1-z)_+ }&=&
{1  \over (1-z)} \left[\ln(z) \ln(1-z)
                                  -{1  \over 2} \ln^2(z)\right]
\\
{\ln(z) \over z}\otimes {\ln(1-z)  \over z }&=&{1  \over 2 z} 
\Bigg[-2 ~{\rm S}_{1,2}(1-z)-
\ln(1-z) \ln^2(z)
\nonumber \\ &&                                    
-2 \ln(z) {\rm Li}_2(1-z)\Bigg]
\\
{\ln(z) \over z}
\otimes \left({\ln(1-z)\over 1-z}\right)_+ &=&{1  \over z} 
                                  \Bigg[{1  \over 2} \ln^2(1-z)
                                  \ln(z)
                           +\ln(1-z) {\rm Li}_2(1-z)
\nonumber \\ &&-{\rm Li}_3(1-z) \Bigg]
\\
{\ln(z) \over z}
\otimes {1  \over (1-z)_+ }&=&{1  \over z} [-{\rm Li}_2(z)+\zeta(2)]
\\
z^2\ln(z)
\otimes \left({\ln(1-z)\over 1-z}\right)_+ &=&z^2 \Bigg[-~{3
\over 4}
                                       -{\rm S}_{1,2}(1-z)+{3  
                                       \over 4  z}
                        +{5  \over 4} \ln(1-z)
 \nonumber \\ &&
                        -{1  \over 4 z^2} \ln(1-z) -{1\over z}
                        \ln(1-z) -{3  \over 4} \ln(z)
 \nonumber \\ &&
                        -{3  \over 2} \ln(1-z) \ln(z) +{1  
                        \over 2} \ln^2(1-z) \ln(z)
 \nonumber \\ &&
                        +{3  \over 4} \ln^2(z)
                        -{1  \over 2} \ln(1-z) \ln^2(z) -{3  
                        \over 2} {\rm Li}_2(1-z) 
 \nonumber \\ &&
                        +\ln(1-z) {\rm Li}_2(1-z)-\ln(z) 
                        {\rm Li}_2(1-z)
 \nonumber \\ &&
-{\rm Li}_3(1-z) \Bigg]
\\
z^2\ln(z)\otimes z^2 \ln(1-z) &=&
-{z^2  \over 2} \Bigg(2~ {\rm S}_{1,2}(1-z)+\ln(1-z) \ln^2(z)
\nonumber \\ &&
                             +2 \ln(z) {\rm Li}_2(1-z)\Bigg)
 \\
z^2\ln(z)
\otimes {1  \over (1-z)_+ }&=&-{1  \over 4}-z+{5  \over 4} z^2-{3  
\over 2} z^2 \ln(z)
\nonumber \\ &&
                      -{z^2  \over 2} \ln^2(z) +z^2 \ln(z) 
                      \ln(1-z)+z^2 {\rm Li}_2(1-z) 
\\
z \ln(z)
\otimes \left({\ln(1-z)\over 1-z}\right)_+\!\! &=&\!\!z 
\Bigg[-{\rm S}_{1,2}(1-z)
                                +\ln(1-z)-{1 \over z}\ln(1-z) 
                                -\ln(z)
 \nonumber \\&&
                            +{1  \over 2} \ln^2(1-z) \ln(z) 
                            -\ln(z) \ln(1-z)+{1  \over 2} 
                            \ln^2(z)
 \nonumber \\&&
                            -{1  \over 2} \ln(1-z) \ln^2(z)
                            -{\rm Li}_2(1-z) +\ln(1-z) 
 \nonumber \\&&
\times {\rm Li}_2(1-z)
                            -\ln(z) {\rm Li}_2(1-z)-{\rm Li}_3(1-z)
                            \Bigg]
 \\
z \ln(z)
\otimes {1  \over (1-z)_+ }&=&
-z \ln(z)-1+z-{z  \over 2} \ln^2(z)+z {\rm Li}_2(1-z)
\nonumber \\&&+z \ln(z) \ln(1-z) 
\\
z \ln(z)\otimes z \ln(1-z) &=&-z~ {\rm S}_{1,2}(1-z)-{z  \over 2} 
\ln^2(z) \ln(1-z)
\nonumber \\ &&
-z \ln(z) {\rm Li}_2(1-z)
\\
\ln(z)
\otimes \left({\ln(1-z) \over 1-z}\right)_+ &=&
-{\rm S}_{1,2}(1-z)+{1  \over 2} \ln^2(1-z)                                           \ln(z)
\nonumber \\ &&
-{1  \over 2} \ln(1-z) \ln^2(z)
                      +\ln(1-z) {\rm Li}_2(1-z)
\nonumber \\ &&
-\ln(z) {\rm Li}_2(1-z)-{\rm Li}_3(1-z) 
\\
\ln(z)\otimes \ln(1-z) &=&-{\rm S}_{1,2}(1-z)-{1  \over 2} 
\ln^2(z) \ln(1-z)
\nonumber \\
&&-\ln(z) {\rm Li}_2(1-z) 
\\
\ln(z)
\otimes {1 \over (1-z)_+ }&=&-{1  \over 2} \ln^2(z)+{\rm Li}_2(1-z)
+\ln(z) \ln(1-z) 
 \\
{1 \over (1-z)_+}
\otimes {1\over (1-z)_+ }&=&2~ {\ln(1-z)  \over (1-z)}-{\ln(z)  
\over (1-z)}-\delta(1-z) \zeta(2) 
 \\
{1 \over z}
\otimes \left({\ln(1-z) \over 1-z}\right)_+ &=&{1  \over 2  z} 
\ln^2(1-z) 
\\
{1 \over z}
\otimes {1  \over (1-z)_+ }&=&{1  \over z} \ln(1-z) 
 \\
z^2
\otimes \left({\ln(1-z)\over 1-z}\right)_+ &=&\left[{1
\over 2}+z-{3  \over 2} z^2\right] \ln(1-z)
                                  -{1  \over 2} z (1-z)
\nonumber \\ &&
                    +{3  \over 2} z^2 \ln^2(z) +{1  \over 2} 
                    z^2 \ln^2(1-z)-z^2 \ln(z) \ln(1-z)
\nonumber \\ &&
                    -z^2 {\rm Li}_2(1-z) 
 \\
z^2
\otimes {1  \over (1-z)_+ }&=&{1  \over 2}+z-{3  \over 2} z^2-z^2 
\ln(z)+z^2 \ln(1-z) 
 \\
z
\otimes \left({\ln(1-z) \over 1-z}\right)_+&=&(1-z) \ln(1-z)+z 
\ln(z)-z \ln(z) \ln(1-z)
\nonumber \\ &&
                  -z {\rm Li}_2(1-z)+{1  \over 2} z \ln^2(1-z) 
 \\
z
\otimes {1  \over (1-z)_+ }&=&1-z+z \ln(1-z)-z \ln(z) 
 \\
1
\otimes \left({\ln(1-z) \over 1-z}\right)_+ &=&-{\rm Li}_2(1-z)
-\ln(z) \ln(1-z)+{1  \over 2} \ln^2(1-z) 
 \\
1
\otimes {1  \over (1-z)_+ }&=&\ln(1-z)-\ln(z) 
\end{eqnarray}
\newpage

\end{document}